\documentclass[lettersize,journal]{IEEEtran}
\usepackage{amsmath,amssymb,amsthm}
\usepackage{amsmath,amsfonts}
\usepackage{algorithmic}
\usepackage{array}
\usepackage[caption=false,font=normalsize,labelfont=sf,textfont=sf]{subfig}
\usepackage{textcomp}
\usepackage{stfloats}
\usepackage{url}
\usepackage{verbatim}
\usepackage{graphicx}
\usepackage{color}
\usepackage{xcolor}
\usepackage{cuted}
\usepackage{bbding}
\usepackage{makecell}
\usepackage{cite}
\usepackage{bm}
\usepackage{algorithm}
\usepackage{setspace}

\makeatletter
\newenvironment{breakablealgorithm}
  {
   \begin{center}
     \refstepcounter{algorithm}
     \hrule height.8pt depth0pt \kern2pt
     \renewcommand{\caption}[2][\relax]{
       {\raggedright\textbf{\ALG@name~\thealgorithm} ##2\par}%
       \ifx\relax##1\relax 
         \addcontentsline{loa}{algorithm}{\protect\numberline{\thealgorithm}##2}%
       \else 
         \addcontentsline{loa}{algorithm}{\protect\numberline{\thealgorithm}##1}%
       \fi
       \kern2pt\hrule\kern2pt
     }
  }{
     \kern2pt\hrule\relax
   \end{center}
  }
\makeatother

\def\BibTeX{{\rm B\kern-.05em{\sc i\kern-.025em b}\kern-.08em
    T\kern-.1667em\lower.7ex\hbox{E}\kern-.125emX}}
\usepackage{balance}
\begin{document}
\title{Secure Communication in the Presence of an RIS-Enhanced Eavesdropper in MIMO Networks}
\author{{Gaoyuan Zhang, Ruisong Si, Boyuan Li*, Zijian Li, Baofeng Ji, Chenqi Zhu, Tony Q.S. Quek, \IEEEmembership{Fellow, IEEE}}
\thanks{Gaoyuan Zhang, Ruisong Si, Baofeng Ji, Chenqi Zhu are with the School of Information Engineering, Henan University of Science and Technology, Luoyang 471023, China. They are also with the Science and Technology Innovation Center of Intelligent system, Longmen Laboratory, Luoyang 471000, China. Gaoyuan Zhang and Baofeng Ji are also with the Institute of Physics, Zhengzhou 451451, Henan Academy of Sciences.}
\thanks{Boyuan Li is with Zhengzhou University, Zhengzhou, China. Zijian Li is with the College of Artificial Intelligence, Dalian Maritime University, Dalian, China. Tony Q. S. Quek is with the Singapore University of Technology and Design, Singapore 487372, and also with the Yonsei Frontier Lab, Yonsei University, Seoul 03722, South Korea.}
\thanks{*Corresponding authors: Boyuan Li (Email: l202311841010602@gs.zzu .edu.cn).}}
\maketitle
\begin{abstract}
In this paper, we pay our attention towards secure and robust communication in the presence of a Reconfigurable Intelligent Surface (RIS)-enhanced mobile eavesdropping attacker in Multiple-Input Multiple-Output (MIMO) wireless networks. Specifically, we first provide a unifying framework that generalizes specific intelligent wiretap model wherein the passive eavesdropper configured with any number of antennas is potentially mobile and 
can actively optimize its received signal strength with the help of RIS by intelligently manipulating wiretap channel characteristics. 
To effectively mitigate this intractable threat, we then propose a novel and lightweight secure communication scheme from the perspective of information theory. The main idea is that the data processing can in some cases be observed as communication channel, and a random bit-flipping scheme is then carefully involved for the legitimate transmitter to minimize the mutual information between the secret message and the passive eavesdropper's received data. The Singular Value Decomposition (SVD)-based precoding strategy is also implemented to optimize power allocation, and thus ensure that the legitimate receiver is not subject to interference from this random bit-flipping. The corresponding results depict that our secure communication scheme is practically desired, which does not require any a prior knowledge of the eavesdropper's full instantaneous Channel State Information (ICSI). Perfect acquisition of ICSI is clearly always not affordable, which is further exacerbated by the RIS involved and the potential mobility of the passive eavesdropper that leads to unavoidable fast fading channels.
Furthermore, we consider the RIS optimization problem from the eavesdropper's perspective, and provide RIS phase shift design solutions under different attacking scenarios. Finally, the optimal detection schemes respectively for the legitimate user and the eavesdropper are provided, and comprehensive simulations are presented to verify our theoretical analysis and show the effectiveness and robustness of our secure communication scheme across a wide range of attacking scenarios.

\end{abstract}


\section{Introduction}\label{section_1}
\IEEEPARstart{T}{h}e issue for achieving Physical Layer Security (PLS) to complemente cryptography-based security protocols at higher layersin 5th-generation (5G) wireless communication network has become increasingly important and continuously received considerable attentions by arena of researchers and entrepreneurs over the last decades \cite{1,3,4}. To address this information-theoretic security requirement, substantial concentrations have been recently achieved on well-designed transmission schemes with the aid of constructive and effective signal processing technology. Among these, Multiple-Input Multiple-Output (MIMO) technology may provide constructive capability most notably including abundant spatial degrees of freedom to confuse the eavesdropper while guaranteeing the communications between legitimate pairs \cite{18,19}. Moreover, an recent emerging technology named Reconfigurable Intelligent Surfaces (RIS) has added a new dimension to PLS design and demonstrated significant potential for augmenting PLS even when the wireless network suffer from a low signal-to-noise ratio (SNR) or strong interference \cite{5,6,R7,q8}. This is mainly attributed to the fact that, by deploying massive low-cost, nearly passive reflecting elements, RIS can dynamically reconfigure the main channel characteristics, effectively converting a ``passive attenuation" environment into a ``programmable gain" medium \cite{100,10}. This reprogramming capability makes the reflected signals constructively added at the legitimate receiver \cite{8,9}, making IRS-enhanced wireless networks as promising solutions to provide resilient, costeffective, and sustainable capabilities for eco-friendly and intelligent 5G and beyond networks \cite{17,OO}.

In the literature, the integration of RIS into MIMO systems seems to be more attractive and imperative for developing security potential. By jointly optimizing the secure beamforming at legitimate user and the reflection beamforming at the RIS, we can achieve constructive signal superposition at the legitimate receiver \cite{12,13,14}. A series of works have demonstrated that such coordinated design not only enhances the main channel capability but also effectively suppresses the interception capability of the eavesdropper to successfully steal information from legitimate users under various antenna architectures \cite{20,21,22}. The reader can refer to \cite{N0} and the references therein for a detailed discussion.
However, a noticeable drawback of this configuration is that it is precisely this unique and powerful programmability that presents a double-edged sword, introduces intractable security threats to critical wireless
communication systems. In fact, it is obviously feasible that eavesdroppers or other adversaries can take advantage of RIS to maliciously enhance eavesdropping performance \cite{23,24,25}, which motivates our imperative study. Thus, unlike most studies that primarily focus on leveraging RIS for legitimate communications, we restrict our attention towards comprehensive
assessment from an offensive perspective, assume that the RIS is maliciously controlled by the passive eavesdropper, and demonstrate
how we can exploit rigorous signal processing algorithms to confuse the eavesdropping while guaranteeing the communications.

\subsection{Motivation and Contribution}\label{subsection1.1}

To the best of our knowledge, a significant body of the recent research has leveraged techniques like RIS and secure beamforming to aid the legitimate communications. However, many secure communication schemes are not applicable to more severe and practical eavesdropping scenarios in the following three aspects. Firstly, unlike the active eavesdropper, the passive eavesdropper cannot jam the conversation between the legitimate transmitter and the legitimate receiver. However, like the legitimate user, the passive eavesdropper can actively optimize its channel characteristics assisted by RIS accounts for better wiretap channel quality.
Secondly, operating in an adversarial environment, the legitimate user cannot obtain cooperation from the eavesdropper, the premise that the perfect achievement of the eavesdropper's full instantaneous Channel State Information (ICSI) is thus difficult and always not affordable in real wiretap scenarios, which practically are typically subject to receiver noise, channel fading, and interference. Thus, ICSI estimation or prediction is complexity intensive and time consuming, which is further exacerbated by the RIS, the potential mobility and passiveness of the eavesdropper. Finally, the premise that the need for the advantage in antenna number is always not easy to satisfy and affordable, most notably the stringent resource
constraints with the legitimate user. Such secure communication scheme, e.g. Artificial Noise (AN) injection, can be completely nullified if the eavesdropper is configured with more antennas than the legitimate user. Motivated by these practical limitations, we propose a robust and lightweight secure communication scheme. The main contributions are summarized as follows.
\begin{itemize}
\item We propose a generalized and intelligent eavesdropping model in MIMO wireless networks, wherein the antenna number of the mobile passive eavesdropper is unlimited, and the mobile eavesdropper can also leverage the controlled RIS to optimize its received signal strength.
\item 
We develop a novel and lightweight anti-eavesdropping scheme. The main idea is that a random bit-flipping approach is carefully involved for the legitimate transmitter to minimize the mutual information (MI) between the secret message and the eavesdropper's received signal. 
\item We propose an Singular Value Decomposition (SVD)-based precoding strategy to optimize power allocation, and correspondingly ensure that the random bit-flipping approach can confuse the eavesdropper but not to interfere with the legitimate receiver, simultaneously enhancing the channel capacity for the primary user. 
\item We construct the optimization problem that the eavesdropper maximizes its received signal power by optimally configuring the phase shifts of its controlled RIS. We then propose the corresponding effective solutions for this eavesdropper-centric RIS optimization.
\item We develop the optimal detection schemes respectively for the legitimate user and the eavesdropper, and conduct comprehensive simulations across a wide range of attacking scenarios. Surprisingly, we observe that our anti-eavesdropping effect is enhanced as the number of Eve's antennas increases, a counterintuitive result that highlights the scheme's unique robustness.
\end{itemize} 

It is worth pointed out that we pay all our attention towards traditional wiretap channel model. The extension to distributed systems \cite{N1} is straightforward but not pursued here. 
Moreover, the ensuing analysis is not tailored uniquely for single eavesdropper, and thus the results may be easily extended to massive eavesdropper case \cite{N2}. Finally, most of the ensuing discussions can directly apply to active eavesdropping attack case \cite{N3}, although it is not conceptually simple. 

\subsection{Paper Organization and Notations}\label{subsection1.2}
The remainder of this work takes the following structure.
Section \ref{section2} reviews the related work. 
Section \ref{section3} introduces the system model. 
Section \ref{section3.4} presents the fundamental principle of our anti-eavesdropping scheme, and Section \ref{section4} develops a novel bit-flipping scheme. 
In Section \ref{section5}, we devise a precoding strategy to ensure reliable reception. 
Section \ref{section6} analyzes the eavesdropper's optimal RIS phase configuration. 
Section \ref{section7} presents the optimal detection schemes for both the legitimate user and the eavesdropper. 
Section \ref{section8} provides comprehensive simulation results. 
Finally, Section \ref{section9} concludes the paper and discusses future research directions.
Notations are summarized in Table \ref{table0}.

  \begin{table*}[htbp]
    \centering
    \small
    \caption{Definition of the Key Mathematical Notations}
    \label{table0}
    \scalebox{0.88}{
    \begin{tabular}{|c|l|c|l|}
      \hline
      \textbf{Symbols} & \textbf{The meaning of these symbols} & \textbf{Symbols} & \textbf{The meaning of these symbols} \\
      \hline
      $M$ & Number of antennas at Alice & $N_e$ & Number of antennas at Eve \\
      $N_b$ & Number of antennas at Bob & $L$ & Number of reflecting elements of the RIS \\
      $\boldsymbol{\Theta}$ & Phase shift configuration matrix of the RIS & $u$ & Transmitted secret message \\
      $\mathbf{U}$ & $M$-dimensional vector replicating $u$ & $\mathbf{S}$ & Intermediate vector after random bit-flipping \\
      $\mathbf{X}$ & Transmit signal vector after precoding $\mathbf{S}$ & $\mathbf{H}_{d}$ & Channel matrix from Alice to Eve (direct path) \\
      $\mathbf{H}_{r}$ & Channel matrix from RIS to Eve & $\mathbf{G}$ & Channel matrix from Alice to RIS \\
      $\mathbf{H}$ & Channel matrix from Alice to Bob & $\mathbf{G}_{E}$ & Equivalent channel matrix observed by Eve \\
      $\mathbf{y}_{B}$ & Signal received by Bob & $\mathbf{y}_{E}$ & Signal received by Eve \\
      $\partial_i$ & Bit-flipping probability: $-1 \to 1$ at $i$-th bit & $\chi_i$ & Bit-flipping probability: $1 \to -1$ at $i$-th bit \\
      \hline
    \end{tabular}}
  \end{table*}

\section{Related work}\label{section2}
\subsubsection{Anti-Wiretapping for MISO System in Distributed WSNs}\label{subsection2.1}
The pioneering theory of PLS was established by Wyner's seminal work on the wiretap channel \cite{R1}. It proved that perfect secrecy is achievable if the main channel's quality surpasses that of the wiretap channel. Motivated by this pioneering work, subsequent research has sought to apply these principles in practical wiretap scenarios. For instance, the authors devised a secure transmission strategy for distributed detection scenarios \cite{R2, R22}. Their scheme exploits random channel variations as an encryption seed to effectively ``blind" the decision fusion center.  For wiretap channels with single antenna, 
\cite{q2} proposes a novel multi-component transmission scheme in the General Multi-Fractional Fourier Transform (GMFRFT) domain, which secures communication by creating asymmetric interference that degrades the eavesdropper's signal quality without sacrificing power on AN.
The work of \cite{q1} proposes a correlation-based secure beamforming that leverages the spatial correlation of the wiretap channel to derive a closed-form secrecy outage probability, thereby enabling more effective maximization of the secrecy rate.
However, the main channel's quality requirements are often difficult and always not affordable to satisfy in real wiretap scenarios \cite{OO}.

\subsubsection{Secure Transmission in Traditional MIMO System}\label{subsection2.2}
The work in \cite{R3} rigorously analyzed the conditions for perfect secrecy in MIMO wiretap channel and precisely characterized the secrecy capacity for arbitrary antenna architectures. Furthermore, \cite{R4} established a computable characterization of the secrecy capacity, formulating it as the saddle-point solution to a minimax problem. This work revealed that an eavesdropper needs a three-to-one antenna superiority to fully compromise the conversation. It also determined that a 2:1 antenna allocation ratio between the legitimate transmitter and legitimate receiver offers optimal resistance against such an attack.
However, MIMO systems lack the adaptability to cope with dynamic and mobile wireless environments and dense multi-user interference.

\subsubsection{Secure Transmission with AN-Aided Jamming}\label{subsection2.3}
A key practical strategy is secure beamforming, which is investigated in \cite{R5}.
This technique maximizes the secrecy rate by employing an Alternating Optimization (AO) algorithm, which decomposes the complex joint optimization problem into two sub-problems that are solved alternately.
However, its efficacy is critically dependent on the availability of the eavesdropper's ICSI. To circumvent this stringent requirement, the concept of AN was introduced. 
As detailed in \cite{q4}, the AN strategy involves injecting a carefully crafted interference signal into the null space of the main channel. This process selectively impairs the eavesdropper's reception while leaving the legitimate user's signal unaffected. Furthermore, based on the proven optimality of the water-filling algorithm, an alternating iterative optimization algorithm is proposed to determine the ideal power ratio between the secret message and the AN to maximize the secrecy capacity. Despite its ingenuity, AN possesses a critical vulnerability  that its security guarantee collapses if the eavesdropper is equipped with more antennas than the legitimate user \cite{R6}.

\subsubsection{RIS-Enhanced MIMO System with Active Eavesdropper}\label{subsection2.4}
The emergence of RIS has added a new dimension to the field of PLS. 
For instance, \cite{R8} proposed an RIS-assisted AN scheme. This scheme improves system secrecy capacity via the alternate optimization of beamforming and the RIS's reflection coefficients.
The work in \cite{q7} introduces a novel reflection modulation scheme, named Superimposed RIS-Phase Modulation (SRPM), which allows RIS to convey additional information by adding phase offsets to its beamforming pattern.
Conversely, the programmability of RIS also introduces new vulnerabilities. In \cite{R9}, researchers conceived of an RIS as a ``green jammer". This device actively reflects signals to create destructive interference to the legitimate receiver. Such an action drastically reduces the main channel's Signal-to-Interference-plus-Noise Ratio (SINR). The work in \cite{q6} further explored the concept of the active eavesdropper, where a distributed neural network, named distributed mixture-of-experts neural network (D-MoENN), was proposed for the detection and localization of such attackers. Furthermore, while a passive eavesdropper only receives signals during the downlink, an active eavesdropper can transmit a spoofing pilot during the uplink to contaminate the channel estimation, thereby ``hijacking" the legitimate transmitter's beamforing for efficient eavesdropping. In \cite{q5}, under a fully reciprocal Time-Division Duplexing (TDD) framework, the authors combined machine learning with the physical characteristics of Massive MIMO. By analyzing only the Received Signal Strength (RSS) without relying on ICSI, they achieved effective detection of active eavesdroppers. These developments signal a clear shift in research towards more intelligent adversarial models.

\subsubsection{Novelty of Our Work}\label{subsection2.5}
Compared with the previous works, the main novelty of this work is that the reliance on the main channel advantage \cite{R1}, antenna number advantage \cite{R6} and the eavesdropper's ICSI \cite{N6} in obtaining secure communication scheme is fully relaxed for defending against RIS-enhanced passive eavesdropper. We propose a new robust and lightweight secure communication scheme, which is practically desired under stringent resource constraints over time-varying mobile channels.
\section{System model}\label{section3}
\begin{figure}[!t]
  \centering
       \includegraphics[width=0.92\linewidth]{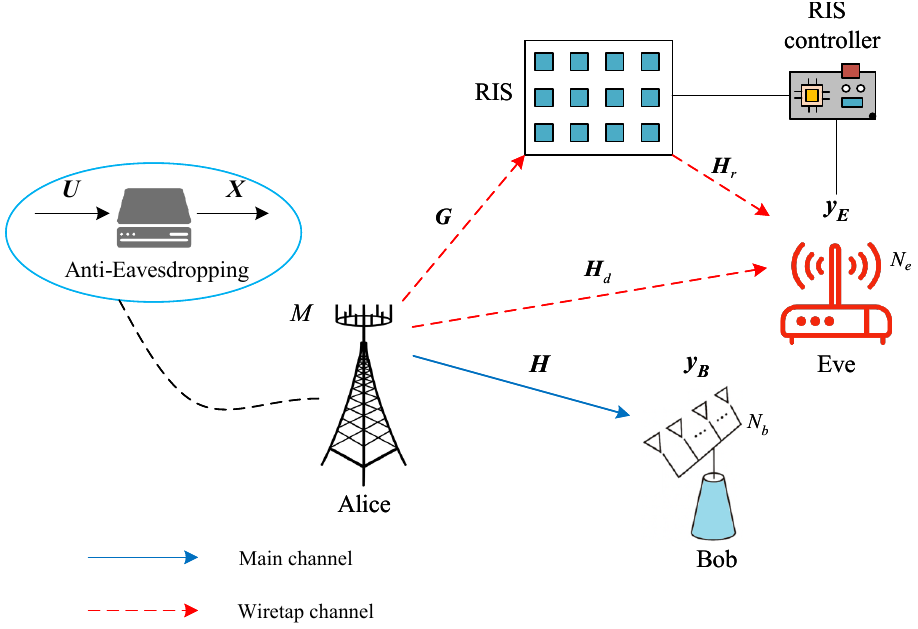}\\
  \caption{MIMO Wireless Networks in the Presence of Eavesdroppers.}\label{tu1}
\end{figure}

\begin{figure*}[!t]
  \centering
   \includegraphics[width=0.9\linewidth]{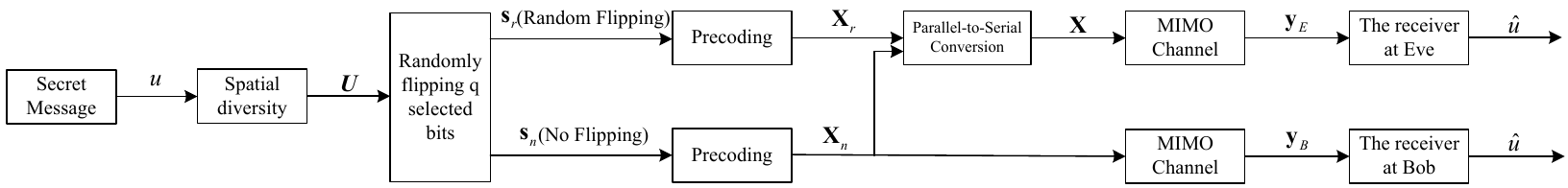}\\
    \caption{The Equivalent Transmission Model.}\label{tu8}
  \end{figure*}

\subsection{System Model}\label{subsection3.1}
As depicted in Fig. \ref{tu1}, we consider a more prevailing wiretap model, which is not studied extensively in the PLS literature but arguably more relevant to practical engineering applications.
In our wiretap model, we consider a MIMO wireless network comprising a legitimate transmitter (Alice) and a legitimate receiver (Bob), operating in the presence of a potential mobile passive eavesdropper (Eve). Alice, Bob, and Eve are configured with $M$, ${N_b}$ and ${N_e}$ antennas, respectively, where ${N_b} \le \frac{M}{2}$. Note here that we do not impose any specification on antenna number of Eve. Furthermore, we assume that there is an RIS composed of $L$ passive reflecting elements, which is designed to enhance passive eavesdropping. 

Without loss of generality and also for analysis convenience and principle clarity, we assume that the secret message to be transmitted is a single bit, denoted as ${u} \in \left\{ {1,-1} \right\}$. To fully utilize spatial diversity, this symbol is replicated to form an $M$-dimensional vector ${\bf{U}} = {\left[ {{u_1}, \cdots ,{u_M}} \right]^T}$.
Subsequently,  this vector  is processed by our well-designed anti-eavesdropping scheme, as illustrated in Fig. \ref{tu8}. This scheme first transforms ${\bf{U}}$  into an intermediate vector
${\bf{S}} = {\left[ {{s_1}, \cdots, {s_M}} \right]^T}$ with the help of the random bit-flipping process. The resulting vector ${\bf{S}}$  is then precoded to generate the final $M$-dimensional transmission signal, denoted as ${{\bf{X}}}  = {\left[ {{x_1}, \cdots, {x_M}} \right]^T}$.
As for the mobile passive eavesdropper, Eve is assumed to control the phase configuration of RIS, represented as $\bf\Theta  = {\mathop{\rm diag}\nolimits} \left( {{{\rm{e}}^{j{\theta _1}}},{{\rm{e}}^{j{\theta _2}}}, \cdots ,{{\rm{e}}^{j{\theta _{\it L}}}}} \right)$, ${\theta _n} \in \left[ {0,2\pi } \right]$, where ${\theta _n}$ denotes the phase shift applied by the $n$-th reflecting element. Due to the severe path loss associated with multiple reflections, we only considers signal paths involving a single reflection, while neglecting second-order and higher-order reflections.
Specifically, Eve manipulates the phase shifts ${\theta _n},n \in \left( {1, \cdots ,L} \right)$, to optimize the passive beamforming gain of the reflected signal at its receiver. for analysis simplification, we assume that the RIS configuration has negligible impact on Bob.

\subsection{Channel Model for the Legitimate User}\label{subsection3.2}
As described above, Bob can only receive the direct signal transmitted from Alice. Consequently, the signal received by Bob can be expressed as:
\begin{eqnarray}\label{math3.1}
  \begin{aligned}
  {{\bf{y}}_{B}} = {\bf{HX}} + {{\bf{n}}_b},
    \end{aligned}
  \end{eqnarray}
where $\mathbf{H}\in {{\mathbb{C}}^{{{N}_{b}}\times \text{M}}}$ denotes the channel fading gain from Alice to Bob, $\mathbf{X}\in {{\mathbb{C}}^{M}}$ represents the transmitted signal from Alice, and ${{\mathbf{n}}_{b}}\in {{\mathbb{C}}^{{{N}_{b}}}}$ is the additive white Gaussian noise (AWGN) vector with zero mean and variance $\sigma _b^2$.
\subsection{Channel Model For Eavesdropper}\label{subsection3.3}
As described in Fig. \ref{tu1}, Eve can receive not only the direct signal transmitted from Alice but also the reflected signal from the RIS under its control. 
The signal received by Eve can be expressed as:
\begin{eqnarray}\label{math3.2}
  \begin{aligned}
  {{\bf{y}}_{{E}}} = {{\bf{G}}_E}{\bf{X}} + {{\bf{n}}_e},
    \end{aligned}
  \end{eqnarray}
where the channel fading gain from Alice to Eve is denoted by ${{\bf{G}}_E} \in {{\mathbb{C}}^{{N_e} \times M}}$, which is given by ${{\bf{G}}_E} = {{\bf{H}}_{{d}}} + {{\bf{H}}_{{r}}}{\bf{\Theta G}}$, where ${{\bf{H}}_{{d}}} \in {{\mathbb{C}}^{{N_e} \times M}}$ represents the direct channel matrix from Alice to Eve, and ${{\bf{H}}_{{r}}}{\bf{\Theta G}}$ denotes the reflected channel matrix from Alice to Eve via the RIS. Specifically, ${{\bf{H}}_{{r}}} \in {{\mathbb{C}}^{{N_e} \times L}}$ is the channel matrix from the RIS to Eve, and ${\bf{G}} \in {{\mathbb{C}}^{L \times M}}$ is the channel matrix from Alice to RIS. The noise vector ${{\bf{n}}_e} \in {{\mathbb{C}}^{{N_e}}}$ consists of AWGN with zero mean and variance $\sigma _e^2$.

Note here that, to reduce the potential probability of being detected, we assume that the passive eavesdropper is mobile. Corresponding, the channel
matrices ${{\bf{H}}_{{d}}}$, ${{\bf{H}}_{{r}}}$ and ${\bf{G}}$ are time-varying.  This renders the previous approach
assuming stationary fading channel statistics impractical and useless. However, the notations have been removed for simplicity. Also, we do not specify each channel model or statistics of ${{\bf{H}}_{{ d}}}$, ${{\bf{H}}_{{r}}}$ and ${\bf{G}}$. This need may arise due to the fact that we aim to provide a unifying framework in treating many of the unreliable communication channels in PLS.
\section{Principle of the Anti-Eavesdropping Scheme}\label{section3.4}

The starting point of our anti-eavesdropping scheme is to nullify Eve's ability to extract any information from the intercepted signals. To accomplish this objective, we devise a two-stage signal processing scheme, which is applied to the confidential data ${\bf{U}}$ at Alice's side before transmission. 

\subsection{Random Bit-flipping}\label{subsection3.1}
The first stage of our scheme utilizes a random bit-flipping process, which is designed from the information-theoretic perspective to minimize the MI between the confidential data $\mathbf{U}$ and the signal intercepted by Eve, ${{\bf{y}}_{{E}}}$. This is achieved by randomly flipping a certain amount bits within confidential data $\mathbf{U}$ prior to transmission. Physically, this corresponds to partitioning Alice's $M$ antennas into two distinct groups: a randomly chosen subset of $q$ antennas is assigned to transmit ``flipped” confidential data, while the remaining $M - q$ antennas transmit the normal (unflipped) and information-bearing confidential data.
\subsection{Null Space Precoding}\label{subsection3.2}
While the random bit-flipping operation effectively confuse Eve to extract useful information, it may also inadvertently interfere with the legitimate receiver Bob. To mitigate this unintended impact, the second stage of our scheme adopts a null-space precoding technique. The implementation of this technique hinges on the construction of a precoding matrix.
Specifically, the ``flipped" confidential data $\mathbf{s}_r$ are projected onto the null space of Bob's channel matrix $\mathbf{H}$, while the ``normal" confidential data $\mathbf{s}_n$ are confined to its range space. This strategic spatial multiplexing guarantees that the legitimate receiver Bob is completely not confusable from the random bit-flipping operation. Conversely, due to the inherent stochastic nature of the channel matrix ${{\bf{G}}_E}$, Eve is unable to perform such a decomposition. It perceives an entangled superposition of signals, which fundamentally prevents the reliable extraction of the secret message.


\section{random bit-flipping approach}\label{section4}
In this section, the implementation of our random bit-flipping approach within the system model depicted in Fig. \ref{tu8} is considered. We give the detailed derivation in the
following.

\subsection{Optimal Condition for Complete Failure of the Wiretap Channel}\label{subsection4.1}
Before launching a compatible random bit-flipping approach for our proposed anti-eavesdropping scheme, it is instructive to point out the condition for complete failure of the wiretap channel from the perspective of information theory. To minimize the information available to the passive eavesdropper's receiver, we construct our objective as minimizing the MI between the secret message $\mathbf{U}$ and Eve's received signal ${{\bf{y}}_{{E}}}$, that is:
 \begin{eqnarray}\label{math4.0}
   \begin{aligned}  
 I\left( {{\bf{U}};{{\bf{y}}_{{E}}}} \right) = h\left( \bf{U} \right) - h\left( {{\bf{U}}\left| {\bf{y}}_E \right.} \right)
\end{aligned}.
 \end{eqnarray}
  According to Data Processing Theorem \cite{O2}, we can easily achieve that
\begin{align}\label{math4.00}
  I\left( \mathbf{U}; \mathbf{y}_E \right) 
  &\le h\left( \mathbf{U} \right) - h\left( \mathbf{U} \mid \mathbf{S} \right) \\
  &= I\left( \mathbf{U}; \mathbf{S} \right).   \notag
  \end{align}



Following the result reported in (\ref{math4.00}), we can easily conclude that if $I\left( {\bf{U};\bf{S}} \right)=0$, then $I\left( {{\bf{U}};{\bf{y}}_{{E}}} \right) \le  0$.  Furthermore, following from  nonnegativity of MI, we have $I\left( {{\bf{U}};{\bf{y}}_{{E}}} \right) \geq  0$. Therefore,  an important conclusion can be intuitively specialized
as that if $I\left( {\bf{U};\bf{S}} \right)=0$, then $I\left( {{\bf{U}};{\bf{y}}_{{E}}} \right) =  0$.
Here, if we regard the data processing from $\mathbf{U}$ to $\bf{S}$ as a $M$th extension of the discrete memoryless channel (DMC) \cite{O2}, then $I\left( {{\bf{U}};{\bf{S}}} \right)$ satisfies: 
  \begin{eqnarray}\label{math4.1}
    \begin{aligned}
    I\left( {{\bf{U}};{\bf{S}}} \right) \le \sum\limits_{i = 1}^M {I\left( {{u_i};{s_i}} \right)} .
      \end{aligned}
    \end{eqnarray}
Following the result reported in (\ref{math4.1}), we can conclude that If $\sum_{i=1}^M I(u_i; s_i) = 0$, then $I(\mathbf{U}; \mathbf{S}) \leq 0$. Since the MI $I\left( {{\bf{U}};{\bf{S}}} \right)$ is non-negative, i. e., $I(\mathbf{U}; \mathbf{S}) \geq 0$, the further conclusion that $I(\mathbf{U}; \mathbf{S}) = 0$ is achieved. Therefore, an instructive conclusion can be intuitively specialized
as that when the MI of the DMC $I\left( {{u_i};{s_i}} \right)=0$, then the MI $I\left( {{\bf{U}};{\bf{y}}_{{E}}} \right) =  0$.
\subsection{Optimal Random Bit-flipping Approach}\label{subsection4.2}
Keep the above observations in mind, we consider the detailed derivation of the random bit-flipping approach.
As to the DMC consisting of the input alphabet ${{u}}$ and output alphabet ${{s}}$, it is characterized by the probability of observing the output symbol ${{s_i}}$
given that we send the symbol ${{u_i}}$. 
The probability transition matrix of the DMC can be given by:
  \begin{multline}\label{math4.10}
    \left[
      \begin{array}{cc}
        1 - \chi_i & \chi_i \\
        \partial_i & 1 - \partial_i
      \end{array}
    \right]
    = \\
    \left[
      \begin{array}{cc}
        P(s_i = -1 \mid u_i = -1) & P(s_i = 1 \mid u_i = -1) \\
        P(s_i = -1 \mid u_i = 1) & P(s_i = 1 \mid u_i = 1)
      \end{array}
    \right].
    \end{multline}
These conditional probabilities ${\partial _i}$ and ${\chi _i}$ can be derived based on the following quantitative analysis. 

A given antenna transmits a random flipped bit only when both of the following conditions are satisfied:

1) The antenna is selected as a randomly flipping antenna, with probability
\begin{eqnarray}\label{math4.3}
  \begin{aligned}
  {P_R} = P\left( {{A_i} = R} \right) = \frac{q}{M};
    \end{aligned}
  \end{eqnarray}

2) The selected flipping antenna successfully flips the bit, with the conditional probabilities given by
\begin{eqnarray}\label{math4.4}
  \begin{aligned}
  \left\{ \begin{array}{l}
P\left( {{s_i} = -1\left| {{u_i} = 1,{A_i} = R} \right.} \right)\\
P\left( {{s_i} = 1\left| {{u_i} = -1,{A_i} = } \right.R} \right).
\end{array} \right. 
    \end{aligned}
  \end{eqnarray}
Combining these two conditions leads to the conditional probabilities ${{\partial }_{i}}$ and ${{\chi }_{i}}$, which are summarized in (\ref{math4.5}) at the bottom of the next page. 
    \begin{figure*}[b]
      \hrulefill
      \begin{eqnarray}\label{math4.5}
      \begin{aligned}
      \left\{
      \begin{array}{rl}
      {\partial _i} &= P\left( {{s_i} = -1 \mid {u} = 1} \right) \\
                    &= P\left( {{s_i} = -1 \mid {u} = 1, {A_i} = R} \right) P\left( {{A_i} = R} \right) + P\left( {{s_i} = -1 \mid {u} = 1, {A_i} = N} \right) P\left( {{A_i} = N} \right) \\
                    &= P\left( {{s_i} = -1 \mid {u} = 1, {A_i} = R} \right) \cdot \frac{q}{M} \\[1.5ex]
      {\chi _i}     &= P\left( {{s_i} = 1 \mid {u} = -1} \right) \\
                    &= P\left( {{s_i} = 1 \mid {u} = -1, {A_i} = R} \right) P\left( {{A_i} = R} \right) + P\left( {{s_i} = 1 \mid {u} = -1, {A_i} = N} \right) P\left( {{A_i} = N} \right) \\
                    &= P\left( {{s_i} = 1 \mid {u} = -1, {A_i} = R} \right) \cdot \frac{q}{M}.
      \end{array}
      \right.
      \end{aligned}
      \end{eqnarray}
      \end{figure*}     
As shown in our previous work \cite{O1}, when the following condition is satisfied:
 \begin{eqnarray}\label{math4.0000}
   \begin{aligned}
 P\left( {{s_i} = 1|{u_i=1}} \right) = P\left( {{s_i} = 1|{u_i=-1}} \right)
\end{aligned},
 \end{eqnarray}
the MI $I\left( {{u_i};{s_i}} \right)=0$. Thus, following the result reported in (\ref{math4.0000}), we have 
that when ${\partial _i} + {\chi _i} = 1$, the channel quality attains nadir from the perspective of the eavesdropper. 
Note here that our random bit-flipping approach does not depend on channel condition, antenna number, or a priori information of the eavesdropper's ICSI.

 \subsection{Optimization of the Random Bit-flipping Fraction}\label{subsection4.3}
 The motivation behind this is to quantify the minimum message manipulation required to achieve perfect secrecy, thereby enhancing the efficiency and practicality of our random bit-flipping approach.
 More specifically, we prove that even with complete knowledge of full system parameters except the random flipping pattern, the eavesdropper is rendered incapable of extracting any information when the proportion of flipped antennas $\frac{q}{M}$ exceeds a certain threshold.

 To satisfy the condition ${\partial _i} + {\chi _i} = 1$, we begin by substituting from (\ref{math4.5}):
  \begin{align}\label{math4.13}
    {\partial _i} + {\chi _i}
    &= \frac{q}{M}\left[ \begin{array}{c}
        P\left( s_i = -1 \mid u = 1, A_i = R \right) \\
     + 
        P\left( s_i = 1 \mid u = -1, A_i = R \right)
      \end{array} \right],
    \end{align}  
and the conditional probabilities satisfy: 
\begin{eqnarray}\label{math4}
  \begin{aligned}
   &P\left( {{s_i} = -1 \mid {u} = 1, {A_i} = R} \right) 
   \\&+ P\left( {{s_i} = 1 \mid {u} = -1, {A_i} = R} \right) 
   &\le 2.
  \end{aligned}
  \end{eqnarray}
 Thus, to ensure that ${\partial _i} + {\chi _i} = 1$, we must have:
 \begin{equation}\label{math4.14}
  \frac{q}{M} \ge \frac{1}{2}.
  \end{equation}
This analysis reveals that the minimum of $q$ is achieved at $q = \frac{M}{2}$. This result signifies that, to guarantee information-theoretic security, at least half of the confidential data $\mathbf{U}$ must be subjected to the random flipping operation. Consequently, the minimum bit-flipping fraction is $\frac{1}{2}$.

\section{Precoding for the legitimate transmitter }\label{section5}

We will develop a precoding strategy to ensure that the random bit-flipping approach given in Section \ref{section4} can confuse the eavesdropper
but not to interfere with the legitimate receiver \cite{N4,R6}. This intractable challenge can be addressed by SVD-based precoding with optimal power allocation such that the legitimate receiver Bob obtains only normal signals $\mathbf{s}_n$, while the eavesdropper Eve receives the whole \(\mathbf{S}\) from all antennas. Note here that we can simultaneously achieving better channel capacity for the primary user. 

First, Alice constructs the composite transmission message \(\mathbf{X}\) by adding the normal information-bearing message \(\mathbf{X}_n\) and the randomly flipped message \(\mathbf{X}_r\):
  \begin{eqnarray}\label{math4.15}
    \begin{aligned}
      \begin{array}{l}
   \mathbf{X} = \mathbf{X}_n + \mathbf{X}_r.
  \end{array}
     \end{aligned}
   \end{eqnarray}
Here, the transmission data \(\mathbf{X}\) is power-normalized such that \(\mathbb{E}\{|\mathbf{X}|^2\} = 1\). The legitimate user's channel matrix \(\mathbf{H}\) undergoes SVD \cite{N5}:
  \begin{eqnarray}\label{math4.16}
    \begin{aligned}
      \begin{array}{l}
        \mathbf{H} = \overline{\mathbf{U}} \mathbf{\Sigma} \mathbf{V}^{\rm H},
  \end{array}
     \end{aligned}
   \end{eqnarray}
where \(\overline{\mathbf{U}} \in \mathbb{C}^{N_b \times N_b}\) is a unitary left singular matrix \((\overline{\mathbf{U}}^{\rm H} \overline{\mathbf{U}} = \mathbf{I})\), \(\mathbf{\Sigma} \in \mathbb{C}^{N_b \times M}\) is a diagonal matrix with singular values on its diagonal, and \(\mathbf{V} \in \mathbb{C}^{M \times M}\) is a unitary right singular matrix \((\mathbf{V}^{\rm H} \mathbf{V} = \mathbf{I})\).
  
  We select the last $M - N_b$ columns of \(\mathbf{V}\) as the null space basis matrix \(\mathbf{Z} \in \mathbb{C}^{M \times (M - N_b)}\), which satisfies \(\mathbf{H} \mathbf{Z} = 0\). The interference message \(\mathbf{s}_r \in \mathbb{C}^{(M - N_b) }\) is precoded through the null space basis:
  \begin{eqnarray}\label{math4.17}
    \begin{aligned}
     \mathbf{X}_r = \mathbf{Z} \mathbf{s}_r. 
     \end{aligned}
   \end{eqnarray}
The normal message \(\mathbf{s}_n \in \mathbb{C}^{N_b }\) is transmitted along the principal subspace of the main channel to maximize received signal power:
\begin{eqnarray}\label{math4.18}
  \begin{aligned}
   \mathbf{X}_n = \mathbf{V}_r \mathbf{\Sigma}_r^{-1} \mathbf{s}_n,
   \end{aligned}
 \end{eqnarray}
  where \(\mathbf{V}_r\) consists of the first \(N_b\) columns of \(\mathbf{V}\), and \(\mathbf{\Sigma}_r^{-1} = \operatorname{diag}(\sigma_1, \ldots, \sigma_r)^{-1}\) normalizes the singular values to ensure proper power allocation, maximizing the Bob's received signal power. Thus, the complete transmitted message is:
  \begin{eqnarray}\label{math4.19}
    \begin{aligned}
     \mathbf{X} &= \mathbf{X}_n + \mathbf{X}_r = \mathbf{V}_r \mathbf{\Sigma}_r^{-1} \mathbf{s}_n + \mathbf{Z} \mathbf{s}_r \\
     &= \left( \mathbf{V}_r \mathbf{\Sigma}_r^{-1}, \mathbf{Z} \right) \begin{pmatrix} \mathbf{s}_n \\ \mathbf{s}_r \end{pmatrix}.
     \end{aligned}
   \end{eqnarray}
 \newpage
  \begin{spacing}{1}
    \begin{breakablealgorithm}
    \caption{Proposed Anti-eavesdropping Scheme.} 
    \label{alg:Framwork} 
    \begin{algorithmic}[1] 
    \REQUIRE ~~\\ 
    $u$: secret message $\in \left\{ { + 1, - 1} \right\}$;\\
    ${M}$: number of transmit antennas;\\
    ${q}$: number of randomly flipped bits.\\
    ${\mathbf{H}}$: Bob's MIMO channel matrix.
    \ENSURE ~~\\
    ${\mathbf{X}}$: the final $M$-dimensional transmit signal vector.
    \STATE \textbf{Initialize }a length-$M$ base signal vector $\mathbf{U}$ by replicating the secret message $u$; 
    \STATE \textbf{Randomly select} $q$ distinct positions in $\mathbf{U}$ to flip, denote them as flipped indices;
    \STATE Identify the remaining indices as non-flipped indices;
    \STATE Flip the bits at the flipped indices to obtain the obfuscation signal component $\mathbf{s}_r$;
    \STATE \textbf{Extract} the non-flipped part of $\mathbf{U}$ as the true information signal component $\mathbf{s}_n$;
    \STATE Perform SVD on Bob's channel $\mathbf{H}$, i.e., $\left[ {\overline {\bf{U}} ,{\bf{\Sigma }},{\bf{V}}} \right] = {\mathop{\rm svd}\nolimits} \left( {\bf{H}} \right)$ (\ref{math4.16});
    \STATE Determine the rank $r$ of the matrix $\mathbf{\Sigma }$;
    \STATE Extract the first $r$ right-singular vectors from $\mathbf{V}$ as the \textbf{principal subspace} $\mathbf{V}_r$;
    \STATE Extract the remaining right-singular vectors from $\mathbf{V}$ as the \textbf{null space }$\mathbf{Z}$;
    \STATE Compute the inverse of the diagonal matrix $\mathbf{\Sigma }(1:r,1:r)$ to obtain the pre-equalization matrix ${\bf{\Sigma }}_r^{{\bf{ - 1}}}$;
    \STATE Construct the final transmit signal vector $\mathbf{X}$ by projecting $\mathbf{s}_n$ into the signal subspace and $\mathbf{s}_r$ into the null space: ${\bf{X}} = {\bf{V}}_r{\bf{\Sigma }}_r^{{\bf{ - 1}}}{{\bf{s}}_n} + {\bf{Z}}{{\bf{s}}_r}$;
    \RETURN ${\bf{X}}$.
    \end{algorithmic}
    \end{breakablealgorithm}
    \end{spacing}
   Here, \(\mathbf{V}_r \in \mathbb{C}^{M \times N_b}\) and \(\mathbf{\Sigma}_r^{-1} \in \mathbb{C}^{N_b \times N_b}\). Correspondingly, Bob's received signal can be given as:
   \begin{eqnarray}\label{math4.201}
     \begin{aligned}
 {{\bf{y}}_B} &= {\bf{HX}} + {{\bf{n}}_b} = {\bf{H}}{{\bf{V}}_{{r}}}{\bf{\Sigma }}_{{r}}^{{\bf{ - 1}}}{{\bf{s}}_n} + {\bf{HZ}}{{\bf{s}}_r} + {{\bf{n}}_b}\\
  &= {\bf{H}}{{\bf{V}}_{{r}}}{\bf{\Sigma }}_{{r}}^{{\bf{ - 1}}}{{\bf{s}}_n} + {{\bf{n}}_b},
   \end{aligned}
    \end{eqnarray}
where the flipped message ${{\bf{s}}_r}$ is completely eliminated by the precoding design, so Bob only receives the normal message \(\mathbf{s}_n\).
  Furthermore, Eve's received signal can be given as:
  \begin{eqnarray}\label{math4.20}
    \begin{aligned}
  {{\bf{y}}_{{E}}} = {{\bf{G}}_E}{\bf{X}} + {{\bf{n}}_e} = {{\bf{G}}_E}{{\bf{V}}_{{r}}}{\bf{\Sigma }}_{{r}}^{{\bf{ - 1}}}{{\bf{s}}_n} + {{\bf{G}}_E}{\bf{Z}}{{\bf{s}}_r} + {{\bf{n}}_e}
  \end{aligned},
   \end{eqnarray}
 Since \(\mathbf{G}_E \mathbf{Z} \neq 0\), the eavesdropper receives both the normal and flipped messages.

In view of the above discussion, we provide the detailed construction process for our proposed anti-eavesdropping scheme ${{\bf{X}}}$ in Algorithm 1.
\section{Received Signal power optimization for the Eavesdropper}\label{section6}

As depicted in Fig. \ref{tu8}, the the legitimate transmitter Alice constructs the transmitted message ${\bf{X}}$ as the sum of the information-bearing message ${{\bf{X}}_{{n}}}$ and a random flipping message ${{\bf{X}}_{{r}}}$. Consequently, the total signal received by the eavesdropper is given in (2). Therefore, the received signal power for Eve is \cite{R9}:
 \begin{eqnarray}\label{math6.2}
  \begin{aligned}
\gamma  = {\left\| {\left( {{{\bf{H}}_{{d}}} + {{\bf{H}}_{{r}}}{\bf{\Theta G}}} \right){\bf{X}}} \right\|^2}
\end{aligned}.
 \end{eqnarray}
  
 \subsection{Problem Formulation}\label{subsection6.1}
 Our objective is to maximize Eve's received signal power as defined in (\ref{math6.2}). Accordingly, the optimization problem can be formulated as:
 \begin{subequations} \label{math6.3}
 \begin{equation}
  \begin{aligned}
P1: {\max _\Theta }\ {\rm{ }}{\left\| {\left( {{{\bf{H}}_{{d}}} + {{\bf{H}}_{{r}}}{\bf{\Theta G}}} \right){\bf{X}}} \right\|^2}
\end{aligned}
 \end{equation}
 \begin{equation}
  \begin{aligned}
{\rm{s}}{\rm{.t}}{\rm{. }}\ 0 \le {\theta _n} \le 2\pi ,{\rm{  }}\forall n = \left\{ {1,2, \cdots ,L} \right\}.
\end{aligned}
 \end{equation}
 \end{subequations}

The objective function in problem (P1) is non-convex. While the phase shifts ${\theta _n}$ are continuous variables, leading to convex constraints, such non-convex optimization problems are inherently difficult to solve to global optimality. These problems are typically nondeterministic polynomial-time hard (NP-hard) and thus difficult to solve directly.

\subsection{Proposed Algorithm }\label{subsection6.2}
To solve Problem (P1), we consider two cases for the transmit message ${\bf{X}}$: known and unknown to Eve. We remind the reader that a performance bound will be achieved in worst case when when ${\bf{X}}$ is known to Eve ${\bf{X}}$. However, our approach differs from \cite{O3} due to our matrix-based channel model.
\subsubsection{Received Power Optimization when ${\bf{X}}$ is Known to Eve}\label{subsubsection6.2.1}
Let ${\bf{v}} = {\left[ {{v_1}, \cdots ,{v_n}, \cdots ,{v_L}} \right]^{\rm{H}}}$, where ${v_n} = {{\rm{e}}^{j{\theta _n}}}$ and $\left| {{v_n}} \right| = 1,\ \forall n = 1, \cdots ,L$. We introduce the transformation: ${{\bf{H}}_{{r}}}{\bf{\Theta GX}} = {{\bf{H}}_{{r}}}{\mathop{\rm diag}\nolimits} \left( {{\bf{GX}}} \right){\bf{v}}$ and define ${\bf{A}} = {{\bf{H}}_{{r}}}{\mathop{\rm diag}\nolimits} \left( {{\bf{GX}}} \right)$. Then the received power at Eve becomes: ${\left\| {\left( {{{\bf{H}}_{{d}}}{\bf{ + }}{{\bf{H}}_{{r}}}{\bf{\Theta G}}} \right){\bf{X}}} \right\|^2} = {\left\| {{{\bf{H}}_{{d}}}{\bf{X + Av}}} \right\|^2}$. Consequently, Problem (P1) can be equivalently formulated as:
\begin{subequations} \label{math6.4}
  \begin{equation}
   \begin{aligned}
 P2: {\max _v}{\rm{  }}\ {\left\| {{{\bf{H}}_{{d}}}{\bf{X + Av}}} \right\|^2}
 \end{aligned}
  \end{equation}
  \begin{equation}
   \begin{aligned}
 {\rm{s.t. }}\ \left| {{v_n}} \right| = 1,\forall n = 1, \cdots ,L.
 \end{aligned}
  \end{equation}
  \end{subequations}
Expanding the objective function Problem (P2):
    \begin{align}\label{math6.5}
      \left\| \mathbf{H}_d \mathbf{X} + \mathbf{A}\mathbf{v} \right\|^2
      &= \left( \mathbf{H}_d \mathbf{X} + \mathbf{A}\mathbf{v} \right)^{\rm H} 
         \left( \mathbf{H}_d \mathbf{X} + \mathbf{A}\mathbf{v} \right) \nonumber \\
      &= \mathbf{v}^{\rm H}  \mathbf{A}^{\rm H}  \mathbf{A}\mathbf{v} 
         + \mathbf{v}^{\rm H}  \mathbf{A}^{\rm H}  \mathbf{H}_d \mathbf{X} \nonumber \\
      &\quad + \mathbf{X}^{\rm H}  \mathbf{H}_d^{\rm H}  \mathbf{A}\mathbf{v} 
         + \left\| \mathbf{H}_d \mathbf{X} \right\|^2.
      \end{align}
since ${\left\| {{{\bf{H}}_{{d}}}{\bf{X}}} \right\|^2}$ is a constant term, Problem (P2) reduces to:
\begin{subequations} \label{math6.6}
  \begin{equation}
   \begin{aligned}
 P2.1: {\max _v}{\rm{  }}\ {{\bf{v}}^{{\rm H} }}{{\bf{A}}^{\rm{H}}}{\bf{Av}} + {{\bf{v}}^{\rm{H}}}{{\bf{A}}^{\rm{H}}}{{\bf{H}}_{{d}}}{\bf{X}} + {{\bf{X}}^{\rm{H}}}{\bf{H}}_{{d}}^{\rm{H}}{\bf{Av}}
 \end{aligned}
  \end{equation}
  \begin{equation}
   \begin{aligned}
 {\rm{s.t.  }}\ \left| {{v_n}} \right| = 1,\forall n = 1, \cdots ,L.
 \end{aligned}
  \end{equation}
  \end{subequations}
Since Problem (P2.1) is still non-convex, we introduce an auxiliary matrix ${\bf{R}}$ and augmented vector $\overline {\bf{v}} $, such that: 
  \begin{eqnarray}\label{math6.7}
    \begin{aligned}
  {\bf{R}} = \left[ \begin{array}{l}
  {{\bf{A}}^{\rm{H}}}{\bf{A}}{\rm{         }}\ \ \ \ \ \ {{\bf{A}}^{\rm{H}}}{{\bf{H}}_{{d}}}{\bf{X}}\\
  {{\bf{X}}^{\rm{H}}}{\bf{H}}_{{d}}^{\rm{H}}{\bf{A }}\ \ \ \ \ \ {\rm{       0}}
  \end{array} \right],\ \overline {\bf{v}}  = \left[ \begin{array}{l}
  {\bf{v}}\\
  1
  \end{array} \right].
     \end{aligned}
    \end{eqnarray}

  Then, the objective Problem (P2.1) becomes: ${\overline {\bf{v}} ^{\rm{H}}}{\bf{R}}\overline {\bf{v}}  = {\rm{tr}}\left( {{\bf{R}}\overline {\bf{v}} {{\overline {\bf{v}} }^{\rm{H}}}} \right)$. Define ${\bf{V}} = \overline {\bf{v}} {\overline {\bf{v}} ^{\rm{H}}} = \left[ \begin{array}{l}
{\bf{v}}{{\bf{v}}^{\rm{H}}}\ {\rm{  }}{\bf{ v}}\\
{{\bf{v}}^{\rm{H}}}\ \ \ {\rm{     1}}
\end{array} \right]$, where ${\bf{V}} \ge 0$ and ${\mathop{\rm rank}\nolimits} \left( {\bf{V}} \right) = 1$. Additionally, since $\left| {{{\bf{V}}_n}} \right| = 1,\ \forall n$, we impose the constraint ${{\bf{V}}_{n,n}} = 1,\forall n$. The rank-one constraint is non-convex, which makes the optimization problem difficult to solve directly. Consequently, we relax this constraint by applying Semidefinite Relaxation (SDR) method.
Then, Problem (P2.1) can be relaxed as:
  \begin{subequations} \label{math6.9}
    \begin{equation}
     \begin{aligned}
   P2.2: {\max _v}\ {\rm{  tr}}\left( {{\bf{RV}}} \right)
   \end{aligned}
    \end{equation}
    \begin{equation}
     \begin{aligned}
    {\rm{s.t.  }}\ {{\bf{V}}_{n,n}} = 1,\forall n = 1, \cdots ,L + 1
   \end{aligned}
    \end{equation}
    \begin{equation}
      \begin{aligned}
     {\bf{V}} \ge 0.
    \end{aligned}
     \end{equation}
    \end{subequations}
Here, we adopt an optimization approach assuming that the transmit message ${\bf{X}}$ is known to Eve. This method enables the derivation of an upper bound on the eavesdropper's received signal power under any circumstances. 

\subsubsection{Received Power Optimization when ${\bf{X}}$ is Unknown to Eve}\label{subsubsection6.2.2}
In this context, the objective is to maximize the combined channel gain ${\left\| {{{\bf{H}}_{{d}}} + {{\bf{H}}_{{r}}}{\bf{\Theta G}}} \right\|^2}$.

Firstly, using the vectorization property of matrix products:
\begin{eqnarray}\label{math6.10}
  \begin{aligned}
    {\mathop{\rm vec}\nolimits} \left( {{\bf{ABC}}} \right) = \left( {{{\bf{C}}^T} \otimes {\bf{A}}} \right) \cdot {\mathop{\rm vec}\nolimits} \left( {\bf{B}} \right)
  \end{aligned},
  \end{eqnarray}
and letting ${\bf{A}} = {{\bf{H}}_{{r}}}$, ${\bf{B}} = {\bf{\Theta }}$, ${\bf{C}} = {\bf{G}}$ we have:
\begin{eqnarray}\label{math6.11}
  \begin{aligned}
    {\mathop{\rm vec}\nolimits} \left( {{{\bf{H}}_{{r}}}{\bf{\Theta G}}} \right) = \left( {{{\bf{G}}^T} \otimes {{\bf{H}}_{{r}}}} \right) \cdot {\mathop{\rm vec}\nolimits} \left( {\bf{\Theta }} \right)
  \end{aligned}.
  \end{eqnarray}
Since ${\bf{\Theta }}$ is a diagonal matrix, its diagonal entries form a vector ${\bf{v}} = {\left[ {{v_1}, \cdots ,{v_L}} \right]^{\rm H}}$, where ${v_n} = {e^{j{\theta _n}}}$. Then:
\begin{eqnarray}\label{math6.12}
  \begin{aligned}
    {\mathop{\rm vec}\nolimits} \left( {\bf{\Theta }} \right) = {\bf{\bar Qv}}
  \end{aligned},
  \end{eqnarray}
  where ${\bf{\bar Q}} \in {\mathbb{C}^{{L^2}{\rm{ \times }}L}}$ is a selection matrix that extracts diagonal entries. Substituting (\ref{math6.12}) into (\ref{math6.12}) gives:
  \begin{eqnarray}\label{math6.13}
    \begin{aligned}
      {\mathop{\rm vec}\nolimits} \left( {{{\bf{H}}_{{r}}}{\bf{\Theta G}}} \right) = \left( {{{\bf{G}}^T} \otimes {{\bf{H}}_{{r}}}} \right){\bf{\bar Qv}}
    \end{aligned}.
    \end{eqnarray}

By the identity ${\left\| {\bf{M}} \right\|^2} = {\left\| {{\mathop{\rm vec}\nolimits} \left( {\bf{M}} \right)} \right\|^2}$, the gain becomes:
  \begin{eqnarray}\label{math6.14}
    \begin{aligned}
\begin{array}{c}
{\left\| {{{\bf{H}}_{{r}}}{\bf{\Theta G}} + {{\bf{H}}_{{d}}}} \right\|^2} = {\left\| {\rm vec\left( {{\bf{H}}_{\it r}{\bf{\Theta G}} + {\bf{H}}_{\it d}} \right)} \right\|^2}\\
 = {\left\| {\left( {{{\bf{G}}^T} \otimes {{\bf{H}}_{{r}}}} \right){\bf{\bar Qv}} + {\mathop{\rm vec}\nolimits} \left( {{{\bf{H}}_{{d}}}} \right)} \right\|^2} = {\left\| {{\bf{Qv}} + {\bf{c}}} \right\|^2},
\end{array}
    \end{aligned}
  \end{eqnarray}
where ${\bf{Q}} = \left( {{{\bf{G}}^T} \otimes {{\bf{H}}_{{r}}}} \right){\bf{\bar Q}}$ and ${\bf{c}} = {\mathop{\rm vec}\nolimits} \left( {{{\bf{H}}_{{d}}}} \right)$. The optimization problem (P1) becomes:
\begin{subequations} \label{math6.15}
  \begin{equation}
   \begin{aligned}
 P3: \begin{array}{l}
{\max _v}\ {\rm{  }}{\left\| {{\bf{Qv}} + {\bf{c}}} \right\|^2} = {\left( {{\bf{Qv}} + {\bf{c}}} \right)^{\rm H}}\left( {{\bf{Qv}} + {\bf{c}}} \right)\\
 = {{\bf{v}}^{\rm H}}{{\bf{Q}}^{\rm H}}{\bf{Qv}} + {{\bf{v}}^{\rm H}}{{\bf{Q}}^{\rm H}}{\bf{c}} + {{\bf{c}}^{\rm H}}{\bf{Qv}} + {\left\| {\bf{c}} \right\|^2}
\end{array}
 \end{aligned}
  \end{equation}
  \begin{equation}
   \begin{aligned}
 {\rm{s.t.  }}\ \left| {{v_n}} \right| = 1,\forall n = 1, \cdots ,L.
 \end{aligned}
  \end{equation}
  \end{subequations}
We reformulate this into a quadratic form using an auxiliary matrix ${\bf{\hat R}}$ and extended vector ${\bf{\hat v}}$, where 
  \begin{eqnarray}\label{math6.16}
    \begin{aligned}
  {\bf{\hat R}} = \left[ \begin{array}{l}
{{\bf{Q}}^{\rm H}}{\bf{Q     }}\ \ {{\bf{Q}}^{\rm H}}{\bf{c}}\\
{{\bf{c}}^{\rm H}}{\bf{Q    }}\ \ \ {\rm{ }}{\left\| {\bf{c}} \right\|^2}
\end{array} \right],{\bf{\hat v}} = \left[ \begin{array}{l}
{\bf{v}}\\
1
\end{array} \right],
\end{aligned}
\end{eqnarray}
the objective Problem (P3) becomes:
\begin{subequations} \label{math6.17}
  \begin{equation}
   \begin{aligned}
 P3.1: \begin{array}{l}{\max _v}\ {\rm{  }}{{\bf{\hat v}}^{\rm{H}}}{\bf{\hat R\hat v}}
\end{array}
 \end{aligned}
  \end{equation}
  \begin{equation}
   \begin{aligned}
 {\rm{s.t.  }}\ \left| {{{\hat v}_n}} \right| = 1,\forall n = 1, \cdots ,{L^2} + 1.
 \end{aligned}
  \end{equation}
  \end{subequations}

Simultaneously, we observe that ${{\bf{\hat v}}^{\rm{H}}}{\bf{\hat R\hat v}} = {\mathop{\rm tr}\nolimits} \left( {{\bf{\hat R\hat v}}{{{\bf{\hat v}}}^{\rm{H}}}} \right)$. Define the lifted variable ${\bf{\hat V}} = {\bf{\hat v}}{{\bf{\hat v}}^{\rm{H}}} = \left[ \begin{array}{l}
{\bf{v}}{{\bf{v}}^{\rm{H}}}\ {\rm{  }}{\bf{ v}}\\
{{\bf{v}}^{\rm{H}}}\ \ \ {\rm{     1}}
\end{array} \right]$ must satisfy ${\bf{\hat V}} \ge 0$, ${\rm{rank}}\left( {{\bf{\hat V}}} \right) = 1$, and ${{\bf{\hat V}}_{n,n}} = 1,\forall n$.  The rank-one constraint renders the problem non-convex. By applying Semidefinite Program (SDP), we relax the rank constraint, obtaining a standard SDP.  The optimization Problem (P3.1) can be reformulated as:
  \begin{subequations} \label{math6.19}
    \begin{equation}
     \begin{aligned}
   P3.2: {\max _v}\ {\rm{ }}{\mathop{\rm tr}\nolimits} \left( {{\bf{\hat R\hat V}}} \right)
   \end{aligned}
    \end{equation}
    \begin{equation}
     \begin{aligned}
   {\rm{s.t.  }}\ {{\bf{\hat V}}_{n,n}} = 1,\forall n = 1, \cdots ,L + 1
   \end{aligned}
    \end{equation}
    \begin{equation}
      \begin{aligned}
     {\bf{\hat V}} \ge 0.
    \end{aligned}
     \end{equation}
    \end{subequations}

    The relaxed Problem (P3.2), as well as Problem (P2.2), are both convex problems and thus can be efficiently solved by off-the-shelf convex solvers like CVX. However, the obtained solution may not satisfy the rank-one constraint, thereby yielding only an upper bound of the original non-convex problem. To recover a feasible rank-one solution, a Gaussian randomization technique can be employed to extract the final beamforming vector ${\Omega ^ * }$.


\section{detection schemes for the Receivers}\label{section7}
This section is dedicated to the detection schemes employed by the legitimate receiver and the eavesdropper.

\subsection{Optimal Detection Scheme for the Legitimate Receiver}\label{subsection7.1}
The detection scheme for the legitimate receiver Bob is given by \cite{O4}:

  \begin{equation}\label{math7.1}
    \begin{aligned}
    \widehat{u} &= 
    \begin{cases}
    1, & \text{if } \Lambda_{\text{opt}} \ge \gamma \\
    -1, & \text{otherwise}.
    \end{cases}
    \end{aligned}
    \end{equation}
Here, $\widehat {{u}}$, ${\Lambda _{\text{opt}}}$ and $\gamma $ denote the detection result, the optimal detection statistic in log-likelihood ratio (LLR) form, and the detection threshold, respectively. The optimal detection statistic ${\Lambda _{\text{opt}}}$ is given by:
    \begin{equation}\label{math7.11111}
      \begin{aligned}
    {\Lambda _{\text{opt}}} &\triangleq \ln \left[ \frac{p\left( {\mathbf{y}_{\mathbf{B}}|u=1} \right)}{p\left( {\mathbf{y}_{\mathbf{B}}|u=-1} \right)} \right] \\
  \end{aligned}.
\end{equation}

  \begin{figure*}[b]
    \hrulefill
    \begin{eqnarray}\label{math7.2}
    \begin{aligned}
      \Lambda_{\text{opt}} 
      &= \ln \frac{
        \sum\limits_{\mathbf{X}} \left[ 
          p(\mathbf{y}_{{B}}|\mathbf{X}) 
          \prod\limits_{i=1}^r P(x_i|u_{i}=1) 
        \right]
      }{
        \sum\limits_{\mathbf{X}} \left[ 
          p(\mathbf{y}_{{B}}|\mathbf{X}) 
          \prod\limits_{i=1}^r P(x_i|u_{i}=-1) 
        \right]
      } 
      \overset{(a)}{=} \ln \frac{
        \sum\limits_{\mathbf{S} \in \mathcal{S}^r} \left[ 
          p(\mathbf{y}_{{B}}|\mathbf{S}) 
          \prod\limits_{i=1}^r P(s_i|u_{i}=1) 
        \right]
      }{
        \sum\limits_{\mathbf{S} \in \mathcal{S}^r} \left[ 
          p(\mathbf{y}_{{B}}|\mathbf{S}) 
          \prod\limits_{i=1}^r P(s_i|u_{i}=-1) 
        \right]
      } \\
      &= \ln \frac{
        \sum\limits_{\mathbf{S} \in \mathcal{S}^r} \left[
          \exp \left( 
            -\frac{ \left\| \mathbf{y}_{{B}} - \mathbf{H} \mathbf{V}_r \Sigma_r^{-1} \mathbf{S} \right\|^2 }{ \sigma_b^2 } 
          \right)
          \prod\limits_{i=1}^r P(s_i|u_{i}=1)
        \right]
      }{
        \sum\limits_{\mathbf{S} \in \mathcal{S}^r} \left[
          \exp \left( 
            -\frac{ \left\| \mathbf{y}_{{B}} - \mathbf{H} \mathbf{V}_r \Sigma_r^{-1} \mathbf{S} \right\|^2 }{ \sigma_b^2 } 
          \right)
          \prod\limits_{i=1}^r P(s_i|u_{i}=-1)
        \right]
      }.
    \end{aligned}
    \end{eqnarray}
  \end{figure*}
 The explicit expression of ${\Lambda _{\text{opt}}}$ is provided at the bottom of this page. Step (a) leverages the Markov chain ${u_i} \to {s_i} \to {x_i}$.  Since the precoder ${\bf{F}}$ projects 
${\bf{S}}$ onto the channel's principal subspace (i.e., ${\bf{X}} = {\bf{Fs}}$), and the degrees of freedom are limited by $r = {\mathop{\rm rank}\nolimits} ({\bf{H}})$, summation over ${\bf{X}}$ is equivalent to summation over ${\bf{S}} \in {S^r}$.
Here, ${\bf{F}} = {{\bf{V}}_r}{\bf{\Sigma }}_r^{ - 1}$, where ${{\bf{V}}_r} \in {\mathbb{C}^{M \times r}}$, ${\bf{\Sigma }}_r^{ - 1} \in {\mathbb{C}^{r \times r}}$. 
An alternative detection statistic with identical performance to (\ref{math7.2}) is provided in the \textbf{Appendix}.

\subsection{Optimal Detection Scheme for the Eavesdropper}\label{subsection7.2}
The detection schemes for the eavesdropper is designed following the same procedure as that for the legitimate user. However, due to differences in the received signals and channel conditions, as well as the exponential growth of ${e^x}$, the explicit expression of its LLR is presented at the bottom of the next page. 
Step (a) is justified by the Markov chain ${u_i} \to {s_i} \to {x_i}$.  Due to our precoding structure, any transmit vector takes the form ${\bf{X}} = {\bf{F}}{{\bf{s}}_n} + {\bf{Z}}{{\bf{s}}_r}$. Consequently, the summation over all possible ${\bf{X}}$ can be equivalently reformulated as a nested summation over ${{\bf{s}}_n} \in {\left\{ {-1,1} \right\}^r}$ and ${{\bf{s}}_r} \in {\left\{ {-1,1} \right\}^{M - r}}$.

  \begin{figure*}[b]
    \hrulefill
    \begin{eqnarray}\label{math7.3}
    \begin{aligned}
      \Lambda_{\text{opt}} 
      &= \ln \frac{
        \sum\limits_{\mathbf{X}} \left[
          p(\mathbf{y}_E|\mathbf{X}) 
          \prod\limits_{i = 1}^M P(x_i|u_{i}=1)
        \right]
      }{
        \sum\limits_{\mathbf{X}} \left[
          p(\mathbf{y}_E|\mathbf{X}) 
          \prod\limits_{i = 1}^M P(x_i|u_{i}=-1)
        \right]
      }                                                                                                       \mathop = \limits^{(a)} \ln \frac{
        \sum\limits_{\mathbf{s}_r \in \{-1,1\}^r,\, \mathbf{s}_n \in \{-1,1\}^{M - r}} \left[
          p(\mathbf{y}_E|\mathbf{s}_r,\mathbf{s}_n) 
          \prod\limits_{i = 1}^M P(s_i|u_{i}=1)
        \right]
      }{
        \sum\limits_{\mathbf{s}_r \in \{-1,1\}^r,\, \mathbf{s}_n \in \{-1,1\}^{M - r}} \left[
          p(\mathbf{y}_E|\mathbf{s}_r,\mathbf{s}_n) 
          \prod\limits_{i = 1}^M P(s_i|u_{i}=-1)
        \right]
      }                                                                                                       
\\[4pt]
      &= \ln \sum\limits_{\mathbf{s}_r \in \{-1,1\}^r,\, \mathbf{s}_n \in \{-1,1\}^{M - r}} 
          \exp \left(
            - \frac{ \left\| \mathbf{y}_E - \mathbf{G}_E \mathbf{V}_r \Sigma_r^{-1} \mathbf{s}_n - \mathbf{G}_E \mathbf{Z} \mathbf{s}_r \right\|^2 }{ \sigma_e^2 } 
            + \sum\limits_{i = 1}^M \ln P(s_i|u_{i}=1)
          \right)                                                                                                           \\[4pt]
      &\quad - \ln \sum\limits_{\mathbf{s}_r \in \{-1,1\}^r,\, \mathbf{s}_n \in \{-1,1\}^{M - r}} 
          \exp \left(
            - \frac{ \left\| \mathbf{y}_E - \mathbf{G}_E \mathbf{V}_r \Sigma_r^{-1} \mathbf{s}_n - \mathbf{G}_E \mathbf{Z} \mathbf{s}_r \right\|^2 }{ \sigma_e^2 } 
            + \sum\limits_{i = 1}^M \ln P(s_i|u_{i}=-1)
          \right).
    \end{aligned}
    \end{eqnarray}
  \end{figure*}

Thus, Eve exhaustively enumerate all possible transmitted  vectors ${\bf{X}}$, and computes the weighted conditional probabilities of the received signal under the two hypotheses $1$ and $-1$, respectively.
Crucially, since the eavesdropper cannot obtain the actual channel flipping probabilities, it adopts an idealized assumption in computing the weighted conditional probabilities: 
the eavesdropper is assumed to know only the overall flipping probability, without knowledge of which specific bits have been flipped.
Taking the logarithm of the weighted sums yields the optimal LLR, enabling binary hypothesis detection.

\begin{table}[t]
  \renewcommand{\arraystretch}{0.9} 
  \small
  \begin{center}
  \caption{Parameters Used in Simulations }\label{table1}
  \scalebox{1.05}{
  \begin{tabular}{cc}
  \hline
  Parameter  & Detailed description \\ \hline
 \begin{tabular}[c]{@{}c@{}}{Number of transmit}\\ {antennas  $M$}\end{tabular} &\begin{tabular}[c]{@{}c@{}}{8 to 16 (default: 9)}\end{tabular}\\  
 \begin{tabular}[c]{@{}c@{}}{Number of RIS}\\ {reflective elements $L$}\end{tabular}  & 9 to 14 (default: 9) \\ \begin{tabular}[c]{@{}c@{}}{Number of legitimate}\\ {user antennas $N_b$}\end{tabular}
 & 2 to 7 (default: 4)\\ \begin{tabular}[c]{@{}c@{}}{Number of eavesdropper}\\ {antennas $N_e$}\end{tabular} &\begin{tabular}[c]{@{}c@{}}{4 and 6 to 16 (default: 4)}\\ {}\end{tabular}\\
 \begin{tabular}[c]{@{}c@{}}{Channel condition }\\{${\bf{H}}$, ${{\bf{H}}_{{d}}}$, ${{\bf{H}}_{{r}}}$ and ${\bf{ G}}$}\end{tabular} &\begin{tabular}[c]{@{}c@{}}{Rayleigh fading with}\\ {normalised average power}\end{tabular}\\ 
  Noise condition ${{\bf{n}}_b}$ and ${{\bf{n}}_e}$ & Complex AWGN\\
  Phase shift of RIS $\bm{\Theta}$ &\begin{tabular}[c]{@{}c@{}}{Random, Optimal, and}\\ {Suboptimal schemes}\\ \end{tabular}\\
  Information sequence length  &200 bits\\
  Number of simulation cycles &\begin{tabular}[c]{@{}c@{}}{Get at least 3000}\\ {frame errors}\end{tabular} \\ \hline
  \end{tabular}}
  \end{center}
 \end{table}

\begin{figure}[ht]
  \centering
       \includegraphics[width=0.77\linewidth]{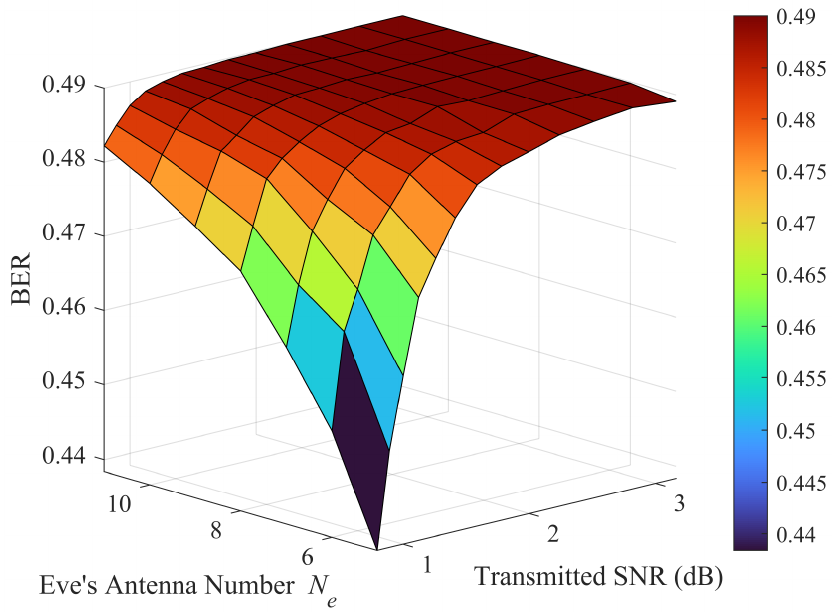}\\
  \caption{Eavesdropper's BER for different numbers of its antennas (${N_e}$) when $M = 9$, ${N_b} = 4$ and $L = 9$.}\label{Ftu1}
\end{figure}
\begin{figure}[ht]
  \centering
       \includegraphics[width=0.77\linewidth]{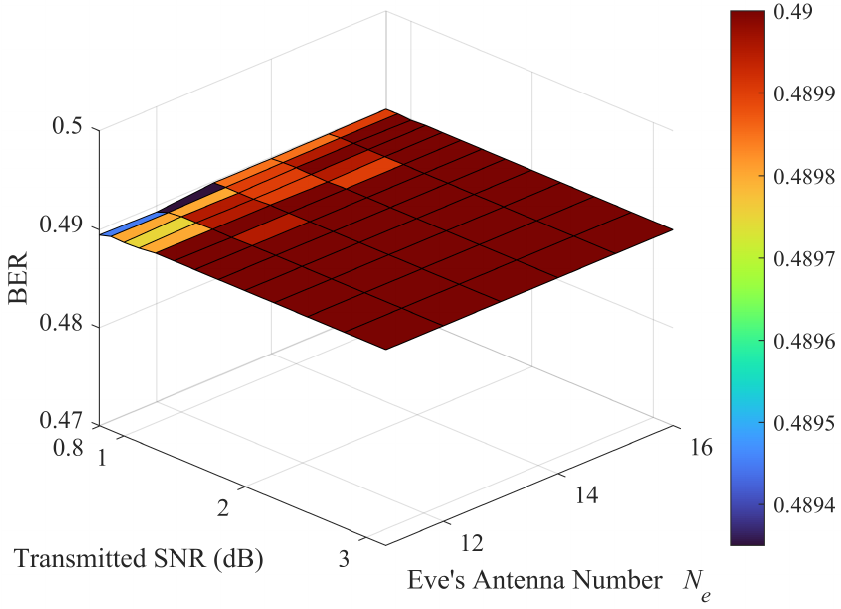}\\
  \caption{Eavesdropper's BER for different numbers of its antennas (${N_e}$) when the number of transmit antennas is fixed at $M = 11$.}\label{Ftu2}
\end{figure}
\begin{figure}[!t]
  \centering
       \includegraphics[width=0.77\linewidth]{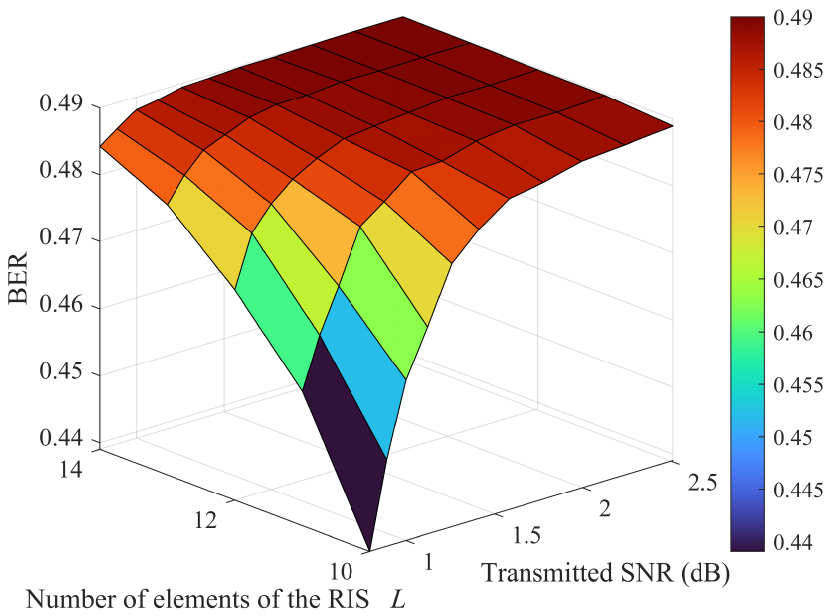}\\
  \caption{Eavesdropper's BER for different numbers of RIS elements ($L$) when $M = 9$, ${N_b} = 4$ and ${N_e} = 4$.}\label{Ftu3}
\end{figure}

\begin{figure}[ht]
  \centering
       \includegraphics[width=0.77\linewidth]{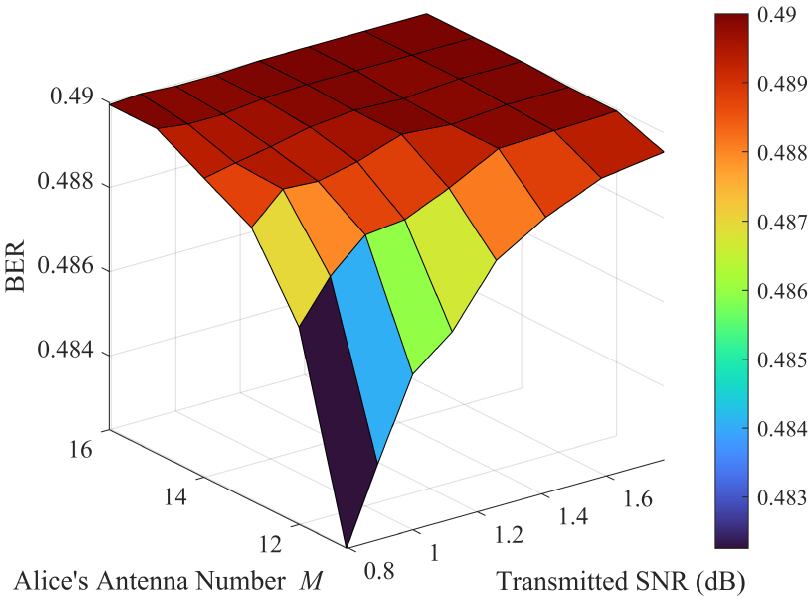}\\
  \caption{Eavesdropper's BER for different numbers of transmit antennas ($M$) when ${N_b} = 4$, $L = 9$ and ${N_e} = 4$.}\label{Ftu4}
\end{figure}

\section{Numerical Results And Discussion }\label{section8}
 We evaluate the secrecy performance of our proposed anti-eavesdropping scheme via Monte Carlo simulations, with the Bit Error Rate (BER) serving as the primary performance metric.
 The system parameters and their respective values used in our simulations are summarized in Table \ref{table1}.
  All wireless channel are modeled as Rayleigh fading channels with normalised average power. It is particularly noteworthy that the channel conditions vary for each bit within each frame. 

\subsection{Detection Performance of the Eavesdropper}\label{subsection8.1}
\begin{figure}[!t]
  \centering
       \includegraphics[width=0.77\linewidth]{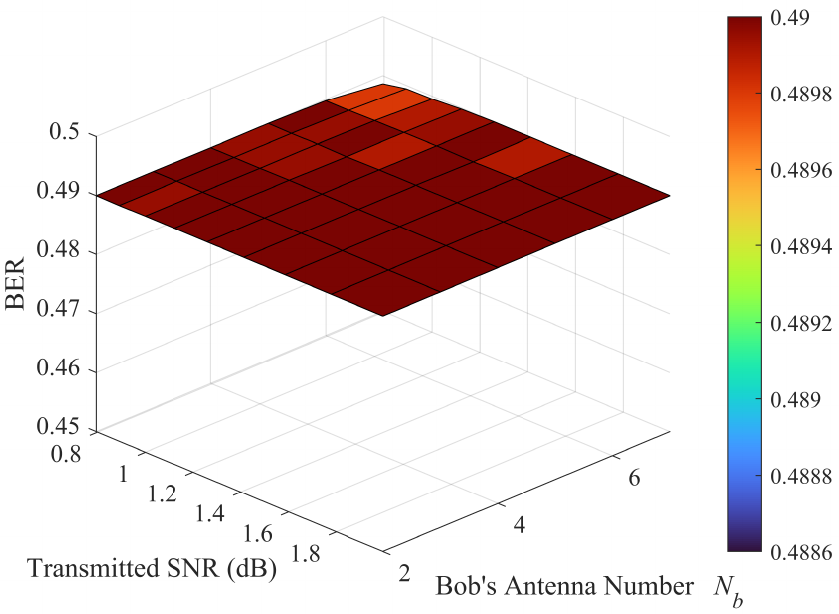}\\
  \caption{ Eavesdropper's BER for different numbers of legitimate user antennas (${N_b}$) when $M = 15$, $L = 9$ and ${N_e} = 8$.}\label{Ftu6}
\end{figure}

Since we concluded in (\ref{math4.14}) that the optimal flipping strategy depends solely on the flipping probability and flipping ratio, and is independent of the specific channel state information, we adopt a unified optimal flipping strategy in the subsequent numerical simulations.  This optimal strategy is implemented through the selected number of flipped bits and the pre-set flipping probabilities that satisfy the condition ${\partial _i} + {\chi _i} = 1$.
As shown in Figs. \ref{Ftu1} to \ref{Ftu6}, we evaluate the BER performance at the Eve under varying transmitted signal-to-noise ratio (SNR) conditions. This analysis is conducted across different configurations of $M$, $N_b$ and $N_e$, as well as different $L$.

As depicted in Figs. \ref{Ftu1} and \ref{Ftu2}, the BER at the Eve gradually increases with the transmitted SNR and eventually approaches 0.49. A closer examination reveals that when the transmitted SNR is set to 0.8, the BER rises significantly as Eve's antenna count increases, due to its improved signal reception. In this case, the proposed scheme effectively interferes with Eve's reception, boosting its BER and enhancing security.

As depicted in Fig. \ref{Ftu2}, the proposed scheme maintains strong anti-eavesdropping performance even when $N_e$ exceeds that of $N_b$. Unlike artificial noise-based methods, its effectiveness is unaffected by $N_e$.

Moreover, experimental results indicate that the scheme can still significantly degrade Eve's decoding performance even when it is equipped with more receiving antennas, thereby enhancing physical layer security under low-SNR conditions.


As shown in Fig. \ref{Ftu3}, the BER at Eve gradually increases with the transmitted SNR and approaches 0.49. Under low-SNR conditions, the BER rises notably as $L$ increases. This is because more RIS elements enhance channel manipulation, improving Eve's ability to recover the signal. Under these conditions, the proposed scheme effectively interferes with Eve's reception, thereby increasing the BER and enhancing the system's physical layer security. Even when Eve controls a greater $L$, the scheme consistently maintains strong anti-eavesdropping performance.


Fig. \ref{Ftu4} demonstrates that increasing $M$ under low-SNR conditions increases Eve's BER and eventually stabilizes it. The reason is that more antennas create additional RIS-assisted paths, improving Eve's signal acquisition.
 However, the proposed scheme leverages this to amplify interference, further degrading Eve's performance and validating  effectiveness in improving physical layer security.



\begin{figure}[!t]
  \centering
       \includegraphics[width=0.77\linewidth]{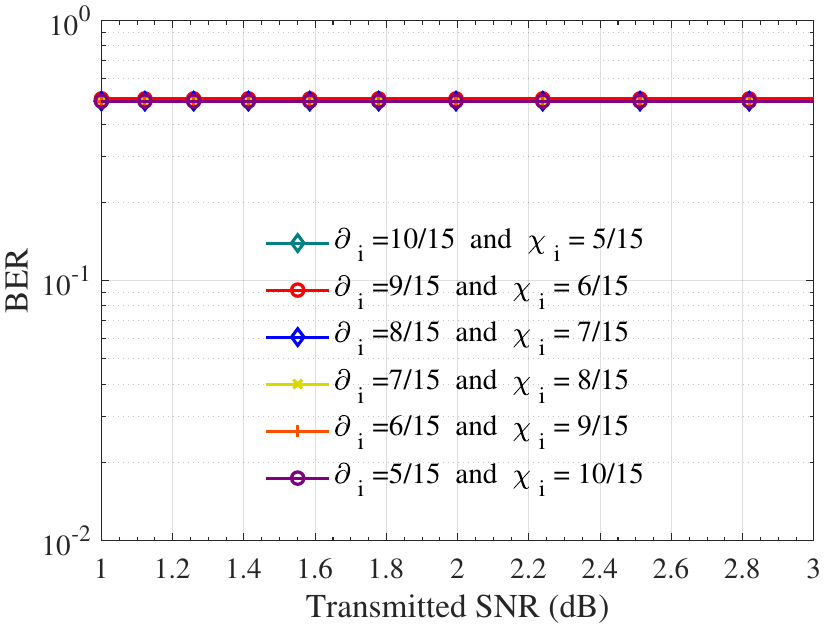}\\
  \caption{Eavesdropper's BER for different flipping probability pairs (${\partial _i}$,${\chi _i}$) when the optimal sum condition ${\partial _i} + {\chi _i} = 1$ is satisfied.}\label{Ftu11}
\end{figure}

Fig. \ref{Ftu6} shows that as transmitted SNR increases, Eve's BER stays stable near 0.49. With fixed transmit and eavesdropper antennas, increasing legitimate user antennas while applying the proposed scheme maintains physical layer security. This limits Eve's ability to recover information and confirms the scheme's robustness across different receiver setups.


\begin{figure}[t]
  \centering
       \includegraphics[width=0.77\linewidth]{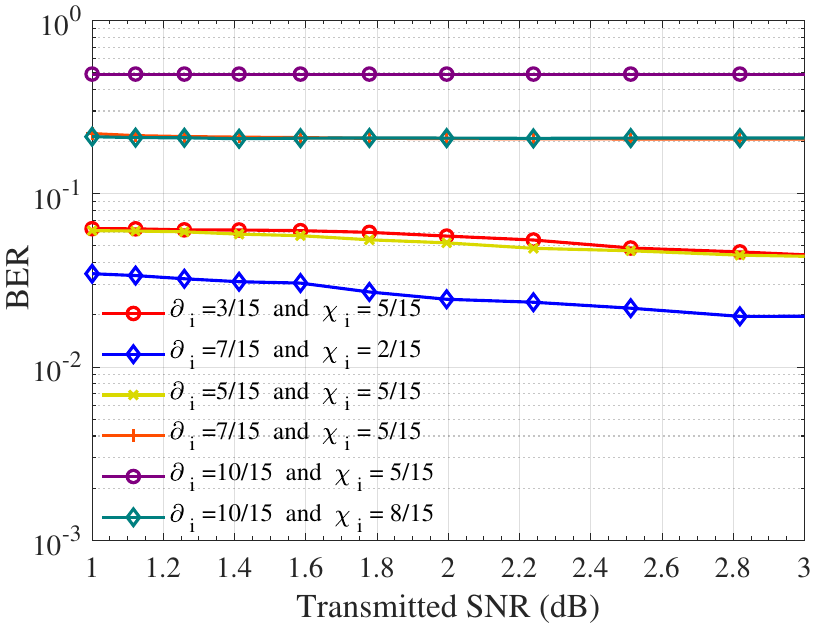}\\
  \caption{Eavesdropper's BER comparison between optimal (${\partial _i} + {\chi _i} = 1$) and suboptimal (${\partial _i} + {\chi _i} \ne 1$) flipping schemes.}\label{Ftu12}
\end{figure}

\begin{figure}[!t]
  \centering
       \includegraphics[width=0.77\linewidth]{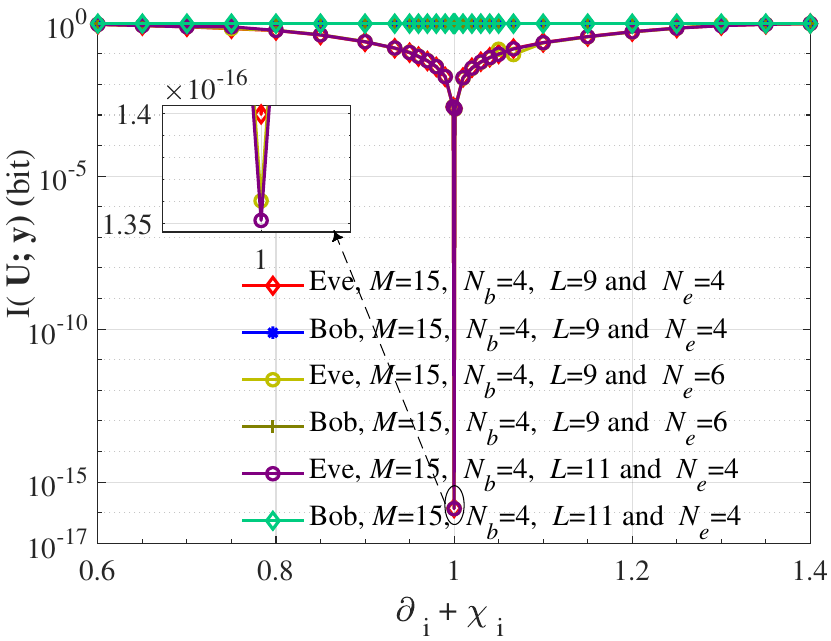}\\
  \caption{Impact of the flipping probability sum ${\partial _i} + {\chi _i}$ on AMI for various system configurations ($L$, $N_e$).}\label{Ftu13}
\end{figure}
\begin{figure}[!t]
  \centering
       \includegraphics[width=0.77\linewidth]{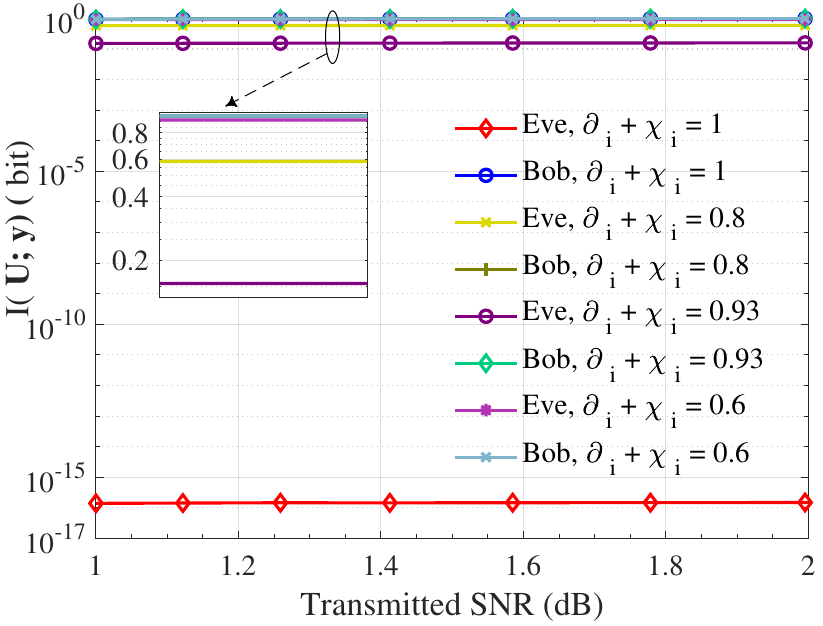}\\
  \caption{Impact of Transmitted SNR on AMI under different flipping probability sum ${\partial _i} + {\chi _i}$.}\label{Ftu14}
\end{figure}

\begin{figure}[!t]
  \centering
       \includegraphics[width=0.77\linewidth]{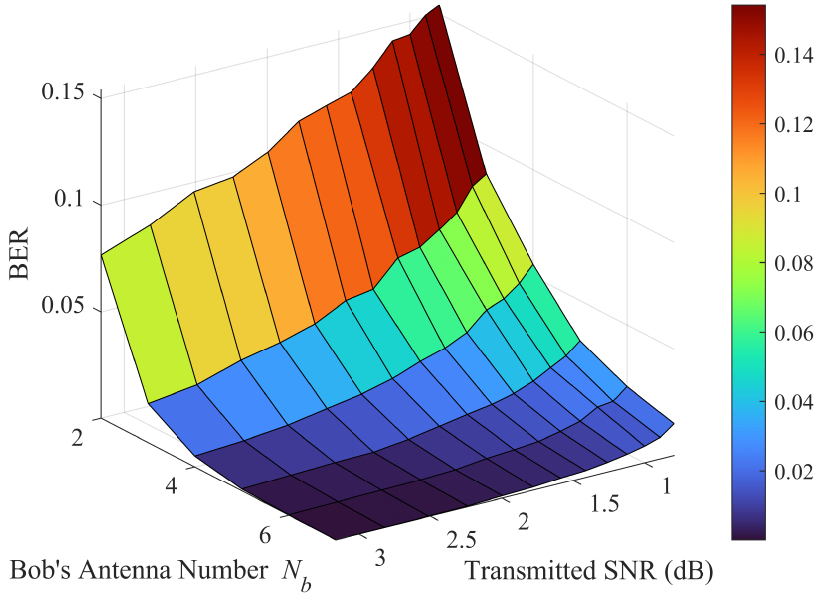}\\
  \caption{Legitimate user's BER for different numbers of its antennas (${N_b}$) when $M = 15$, $L = 9$ and  ${N_e} = 4$.}\label{Ftu7}
\end{figure}

\begin{figure}[!t]
  \centering
       \includegraphics[width=0.77\linewidth]{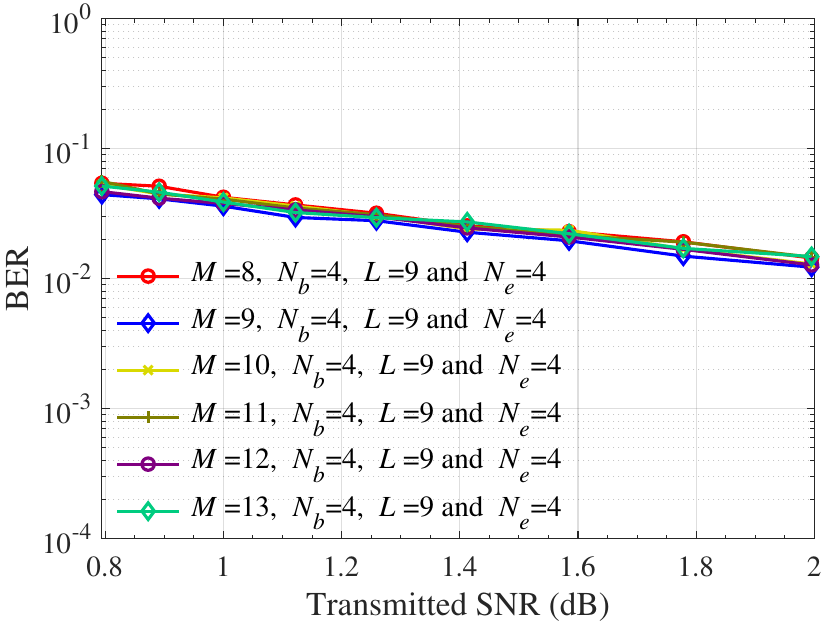}\\
  \caption{Legitimate user's BER for different numbers of transmit antennas ($M$) when ${N_b} = 4$, $L = 9$ and  ${N_e} = 4$.}\label{Ftu8}
\end{figure}

Figs. \ref{Ftu11} and \ref{Ftu12} evaluate the BER performance at the eavesdropper under varying transmitted SNR conditions. The analysis is conducted under the setting $M=15$, $N_b=4$, $N_e=4$ and $L=9$.
Fig. \ref{Ftu11} reveals that as long as the optimal flipping condition ${\partial _i} + {\chi _i} = 1$ is met, Eve's BER consistently converges to 0.5, regardless of the specific values of the flipping probabilities, ${\partial _i}$ and ${\chi _i}$.  With all other system parameters held constant, applying different probability pairs such as ${\partial _i} = 0.6$, ${\chi _i} = 0.4$ or ${\partial _i} = 0.33$, ${\chi _i} = 0.67$ yields nearly identical security performance. This demonstrates that our scheme's effectiveness is predicated on the sum condition rather than the individual probability values, confirming the flexibility and robustness of our proposed optimal strategy.

In contrast, Fig. \ref{Ftu12} shows when the flipping probabilities fail to satisfy the optimal condition ${\partial _i} + {\chi _i} = 1$, the secrecy performance degrades significantly. With a suboptimal setting, such as ${\partial _i} + {\chi _i} < 1$,  Eve's BER drops considerably below 0.5 and continues to decrease as the SNR increases. This indicates information leakage to the eavesdropper and confirms that adherence to our derived optimal condition is critical for ensuring the effectiveness of the anti-eavesdropping scheme.

As shown in Figs. \ref{Ftu13} and \ref{Ftu14}, we evaluate the MI between the original transmitted message $\mathbf{U}$ and the received signals at both Bob and Eve under different bit-flipping probability conditions.
Fig. \ref{Ftu13} demonstrates the presence of a security null at the optimal flipping probability (${\partial _i} + {\chi _i} = 1$), corresponding to the point at which Eve's MI reaches its minimum. At this optimal point, the eavesdropper is effectively unable to extract any meaningful information. The results further confirm that as the flipping probability approaches this optimal value, the MI for Eve consistently decreases, reinforcing the effectiveness of the proposed scheme in degrading eavesdropping performance.
The depth of this null is influenced by system parameters: more RIS elements or eavesdropper antennas deepen the null, enhancing physical layer security. These trends align with Figs. \ref{Ftu2} and \ref{Ftu3}, confirming the effectiveness of the proposed anti-eavesdropping scheme. Meanwhile, Bob's MI remains stable near 1, demonstrating that the precoding reliably preserves legitimate communication.


Fig. \ref{Ftu14} shows the variation of MI for Bob and Eve with transmitted SNR under different flipping probabilities. As the SNR increases, the resulting MI tends to stabilize. Moreover, flipping probabilities closer to this optimal value consistently lead to lower MI for Eve, further demonstrating the robustness of the proposed scheme. 

\subsection{Detection performance of the Legitimate User}\label{subsection8.2}
In Figs. \ref{Ftu7} to \ref{Ftu8}, we evaluate the BER performance at Bob under various transmitted SNR conditions. This analysis considers different configurations of $M$ and $N_b$.

The analysis shows that Bob consistently maintains a lower BER compared to Eve, indicating that the precoding design effectively protects Bob from the impact of the flipping strategy. For example, in Fig. \ref{Ftu7}, when the transmitted SNR is 0.8, Bob's BER is approximately $4 \times {10^{ - 2}}$, and it decreases as the SNR increases.

As shown in Fig. \ref{Ftu8}, the BER at Bob decreases as the transmitted SNR increases. Further analysis reveals that under varying SNR conditions, changes in $M$ have a negligible influence on BER when $N_b$ remains fixed — the BER remains relatively stable. This can be attributed to the fact that $N_b$ determines the number of signal branches unaffected by random flipping, which primarily governs the reception performance. As the SNR increases, Bob's ability to recover the secret message improves due to enhanced reception of the effective signal. The robustness of the proposed scheme is further supported by the simulation results: even under different transmit antenna configurations, the system consistently maintains a low BER provided that $N_b$ remains unchanged, thereby ensuring reliable communication.

As illustrated in Fig. \ref{Ftu7}, further analysis shows that increasing $N_b$ significantly reduces BER performance under low-SNR conditions.
This is because both higher SNR and more receive antennas enhance Bob's ability to capture signals and accurately recover the secret message. These results further validate the effectiveness of the proposed scheme, demonstrating that reliable communication and low BER can be maintained even under challenging low-SNR scenarios.

\subsection{The impact of phase optimization on the eavesdropper's received signal power}\label{subsection8.3}
\begin{figure}[t]
  \centering
       \includegraphics[width=0.77\linewidth]{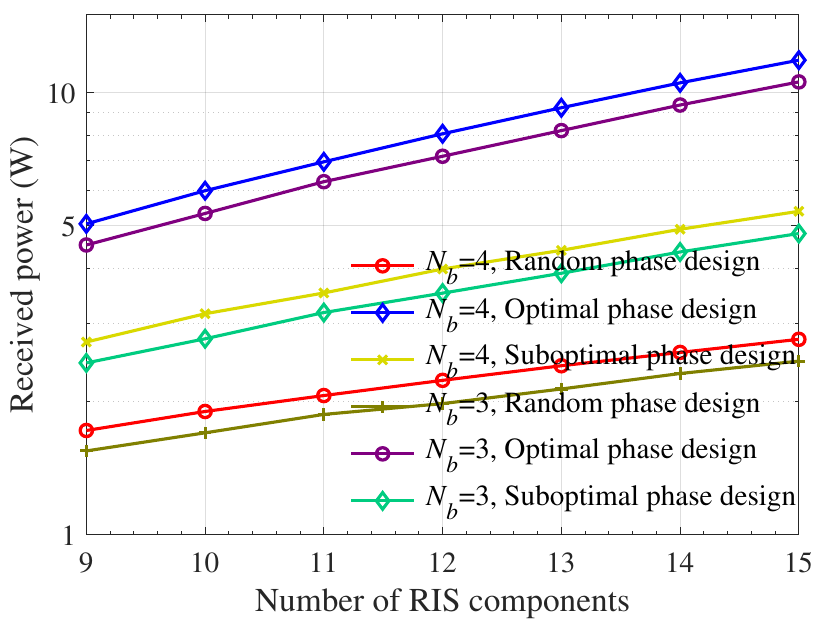}\\
  \caption{Eavesdropper's Receive Power under Different RIS Phase Designs and Legitimate User Antennas (${N_b}$)}\label{Ftu9}
\end{figure}

Fig. \ref{Ftu9} 
illustrate how the received power and corresponding optimization gains vary with $L$ under different strategies. When Eve has knowledge of the transmitted signal, the optimal phase design achieves the highest received power more than three times that of a random configuration and the gain increases as $L$ grows. In contrast, when the transmitted signal is unknown, a suboptimal strategy yields lower power about 1.5 times that of the random case but still improves with more RIS elements. Moreover, the received power also grows with $N_b$, owing to the employed precoding design.

\begin{figure}[t]
  \centering
       \includegraphics[width=0.77\linewidth]{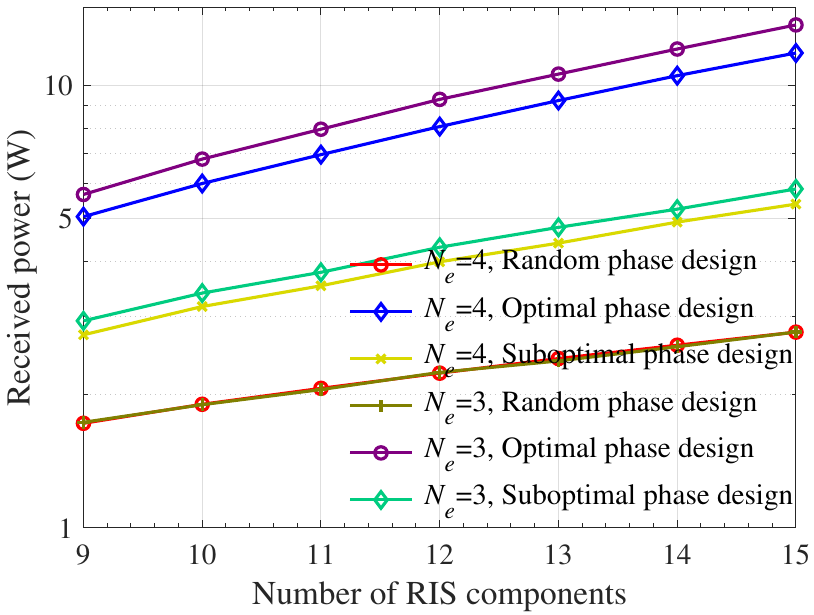}\\
  \caption{Eavesdropper's Received Power under Different RIS Phase Designs and Eavesdropper Antennas (${N_e}$)}\label{Ftu15}
\end{figure}

Fig. \ref{Ftu15} illustrates the per-antenna received power as a function of $L$ under different phase design strategies. As expected, the optimal design significantly outperforms the suboptimal and random schemes, and the received power increases with the number of RIS elements for all approaches.
An interesting observation is that, under optimized designs, increasing $N_e$ from 3 to 4 paradoxically leads to a reduction in per-antenna received power. This is likely due to the total harvested energy being spread over a larger array. In contrast, the power achieved with the random phase design remains largely unaffected by the number of antennas.


Given that the optimal strategy provides the highest possible received power, it represents the theoretical upper bound of the eavesdropper's capability. Therefore, to rigorously evaluate the robustness of our proposed anti-eavesdropping techniques, the performance evaluation presented above was conducted under this worst-case attack scenario.



\section{Conclusions and Future Work}\label{section9}
We develop a novel anti-eavesdropping scheme, based on random bit-flipping and precoding design, to secure  MIMO communications against a passive, RIS-enhanced eavesdropper. 
This anti-eavesdropping framework can provide a promising and potent avenue for defending next-generation wireless networks against intelligent and adaptive security threats.
For future work, several promising research directions can be pursued. A primary direction involves extending our framework to more challenging multi-eavesdropper environments, where multiple malicious nodes may collaborate to intercept the secret message \cite{O5}. Such a scenario necessitates the development of more sophisticated precoding designs capable of simultaneously neutralizing multiple, spatially threats. Furthermore, the joint optimization of security and resource efficiency presents another compelling avenue of research \cite{R8}. 
This entails formulating multi-objective optimization frameworks to balance security and efficiency, thereby facilitating green and secure next-generation communication systems.

\section*{Appendix}\label{Appendix}
We provide an alternative detection statistics that achieves the same performance as the one shown in equation (\ref{math7.2}):
\begin{equation}\label{mathappendix1}
  \begin{aligned}
  \Lambda_{\mathrm{opt}} 
 = \ln \frac{\sum\limits_{\mathbf{S} \in \mathcal{S}^r} \left[ \exp \left( -\frac{\left\| (\mathbf{\overline{{U}}}^{\rm{H}}\mathbf{y}_B)_{1:r} - \mathbf{S} \right\|^2}{\sigma_b^2} \right)\prod\limits_{i = 1}^r P(s_i|u_{i}=1) \right]}{\sum\limits_{\mathbf{S} \in \mathcal{S}^r} \left[ \exp \left( -\frac{\left\| (\mathbf{\overline{{U}}}^{\rm{H}}\mathbf{y}_B)_{1:r} - \mathbf{S} \right\|^2}{\sigma_b^2} \right)\prod\limits_{i = 1}^r P(s_i|u_{i}=-1) \right]}.
  \end{aligned}
  \end{equation}
  


\end{document}